\documentclass[aps,twocolumn,amsmath,amssymb]{revtex4-1}
\usepackage{xcolor}
\usepackage{graphicx} %including figures
\usepackage[caption=false]{subfig}

\usepackage{gensymb} 
\usepackage{amsmath} 
\usepackage{braket} 
\usepackage{booktabs} 
\usepackage[colorlinks=true]{hyperref} %color in references to fig,bib,etc
\usepackage{ulem} %for \sout{} and \uline{}

\hypersetup{colorlinks,
	citecolor=blue
}
\usepackage[capitalize]{cleveref}

\newcommand{\ba}{\begin{eqnarray}}
\newcommand{\ea}{\end{eqnarray}}

\newcommand{\non}{\nonumber}

\begin{document}

%\preprint{} 

%\title{Preliminary: Development of a new superscaling $\Delta$-resonance model and extension of the SuSAv2-inelastic approach to the neutrino sector}
\title{
%Charged-current neutrino-induced single-pion production in superscaling and relativistic distorted wave impulse approximation models (provisional title)\\
%\ttv{
Charged-current neutrino-induced single-pion production in the superscaling approach and the relativistic distorted-wave impulse approximation}
%}


\author{J.~Gonzalez-Rosa$^a$}
\author{A.~Nikolakopoulos$^b$}
\author{M.~B.~Barbaro$^c$}
\author{J.~A.~Caballero$^a$}
\author{R.~Gonz\'alez-Jim\'enez$^a$}
\author{G.~D.~Megias$^a$}
\email{Contact author: megias@us.es}
\affiliation{$^a$Departamento de F\'isica At\'omica, Molecular y Nuclear, Universidad de Sevilla, 41080 Sevilla, Spain}
\affiliation{$^b$Physics Department, University of Washington, Seattle WA}
\affiliation{$^c$INFN, Sezione di Torino, Via P. Giuria 1, 10125 Torino, Italy}

%\email[]{}
%\thanks{}


\date{\today}

\begin{abstract}  
		In this work, we present a detailed comparison of the SuSAv2 (SuperScaling Approach version 2) and RDWIA (Relativistic Distorted-Wave Impulse Approximation) models with measurements of charged-current neutrino-induced single-pion production from different experiments (T2K, MINERvA and MiniBooNE), studying the differences between the two theoretical descriptions. The neutrino energy range in these experiments spans from hundreds of MeV to roughly 20 GeV,
        %A.N: I changed tens to hundreds because any energy below m_\mu + m_\pi ~ 250 MeV is irrelevant for CC pion production
        %\sout{, peaking respectively at different energies},
        and the nuclear targets are mainly composed of $^{12}$C.
        %\sout{ or $^{40}$Ar}. 
        %\sout{For the description of the single-nucleon inelastic structure functions within the SuSAv2 approach, we make use of the ANL-Osaka DCC model,} 
        The SuSAv2 model uses the single-nucleon inelastic structure functions from the ANL-Osaka DCC model, which allows for a separation of pion production channels, distinguishing between the $\pi^+$, $\pi^-$ and $\pi^0$ final states. In the RDWIA approach, the Hybrid model developed by the Ghent group is used for the description of the boson-pion-nucleon vertex. 
        %{\ttblue (We could say something about the findings/results.)}{\ttgreen Fine to me}
        % this which is applicable to the entire energy range of interest for accelerator-based neutrino-oscillation experiments.
        %{\ttorange [Something else about RMF/RDWIA?]}
\end{abstract}
%\keywords{}

\maketitle
% must follow title, authors, abstract, and keywords
%\maketitlesync

%{\ttblue Raul: ttblue.}
%{\ttorange Guille: orange}
%\textcolor{red}{Jesús: red or textcolor red}
%{\ttv Mari: violet}
%{\ttgreen JA: eliminate colors if clear enough. Only maintain those comments that still need to be addressed} {\ttorange Done.}
\section{Introduction}\label{Introduction}
Neutrino oscillation experiments constitute a fundamental tool for exploring the properties of neutrinos, such as the possible violation of charge-parity symmetry or their mass hierarchy \cite{taroni_nobel_2015,abe_constraint_2020,T2K:2025wet}.
The interaction of the neutrino with the nucleus is a key piece in the reconstruction of the neutrino energy in oscillation experiments, a fundamental step in the determination of neutrino oscillation parameters and a source of major uncertainties in the analysis. As such, multiple theoretical and experimental efforts have been developed for this purpose.
% understanding the weak particle interaction with the nucleus. In these 
In these experiments, multiple channels are involved. For example, in the neutrino energy range of hundreds of MeV to a few GeV, the dominant interaction of the neutrino with the nucleus is the quasielastic (QE) process in which a single nucleon is knocked-out from the nucleus. 
%This is known as a quasielastic (QE) process. 
In spite of this process being heavily relevant in the few GeV region, in experimental inclusive 
%{\ttblue (needs define `inclusive', isn't it?)} {\ttorange (and be careful about inclusive definition in experimental context and our inclusive vs. semi-inclusive definition)} 
analyses, where all reaction channels are considered, other processes also give crucial contributions to the cross section, like two-particle two-hole interaction, the excitation of the nucleonic resonances, or deep-inelastic scattering (DIS).  
%{\ttv What is the difference between "excitation of the nucleonic resonances" and "resonance processes"?}. {\ttorange I think it is a typo or maybe the last process was supposed to be DIS. I have rewritten it as DIS.} 
Hence, a reduction in the nuclear-medium uncertainties associated with all these processes, including neutrino-induced pion production, %\sout{resonance processes}
is needed to improve cross-section measurements and neutrino oscillation analyses. %\sout {to determine the weak interaction properties.} 
%{\ttblue (About the last couple of sentences: I would remove 'resonance processes' and simply say 'pion production'. Nucleonic resonance production is just one contribution to pion production, one cannot separate experimentally resonance pion production from background pion production, only with a model. So, talking about resonance channel is a bit misleading unless you clearly define what you mea-n.)} {\ttorange I would also rephrase "weak interaction properties". The relevant uncertainties in oscillation experiments are in the description of the nuclear properties and the nuclear-medium effects, not in the weak interaction properties.}

 Most experiments -- MiniBooNE \cite{aguilar-arevalo_MiniBooNE_2004}, MicroBooNE \cite{MicrobooneOnline}, T2K \cite{T2KOnline}, MINERvA\cite{MINERvAOnline}, NO$\nu$A \cite{NOvAOnline} and future ones like HyperK \cite{Hyper-Kamiokande:2018ofw} or DUNE \cite{Masud2019} -- operate on different energy ranges, such as 0.5-1 GeV (MicroBooNE, T2K, MiniBooNE) or around 10 GeV (DUNE, ArgoNEUT\cite{soderberg_argoneut_2009}). 
 The importance of each channel depends on the energy in which the experiment operates: quasielastic scattering dominates at a neutrino energy of $0.5-1$ GeV, while pion production 
 %\sout{from resonance production} 
 and other inelastic processes become dominant at higher energies. 
 Diverse experimental efforts have been devoted to the study of these different reaction mechanisms. This is the case of 
 %of the interaction of different channels. 
 %One of these efforts is the study of 
 charged-current neutrino interactions with only one pion detected in the final state of the process, together with the scattered lepton. 
 These so-called CC1$\pi$ %\sout{processes} 
 events can be modeled as the excitation of a nucleonic resonance followed by its decay into nucleons, pions and other mesons, non-resonant pion production processes
%\sout{, contributions from resonances other than the $\Delta$ one,}{\ttblue (first time delta is mentioned, not good sentence for that)} 
or 
%\sout{pion-exchange} 
intranuclear pion rescattering effects. %{\ttv what does this mean exactly?}. 
 %\sout{In other words, in these experimental data, the charged-current pion-production resonance channel is included, but the data are not uniquely limited to it.}{\ttblue (The data don't know about resonances, cascade FSI or non-resonant contributions, those are theoretical models. Data are defined by the signal definition.)} 
 %\sout{These studies allow us to corroborate and compare the theoretical predictions made from pion production models.}{\ttorange: Does the previous sentence want to state something like:}  
 These studies also allow us to compare and validate theoretical models for the description of neutrino-induced pion production. 
 %{\ttorange Then, I would keep something like the new sentence.}

%\textcolor{red}{I need to add new up-to-date developments.}
 
On the theoretical 
%\sout{front} 
side, several groups \cite{REIN198179,fogli_new_1979,PhysRevD.74.014009,schreiner__1973,
matsui_quark-hadron_2005,PhysRevD.87.113009, PhysRevC.90.025501,YAO2019109,SOGARWAL2022122494,PhysRevC.109.024608,Yan2024} have
studied neutrino-induced pion production on nuclei, %{\ttblue (I added 'neutrino-induced because the history of pion production models is long and most of the models are not based on Fermi gas approaches, that's something specific from our neutrino community)}
providing different descriptions of the initial nuclear state, the pion production from a bound nucleon, and the subsequent pion-nucleon interaction within the residual nucleus. Most of the initial studies were based on  Fermi gas approaches of non-interacting nucleons~\cite{PhysRevC.48.3078,alberico_notitle_1997}, but recently more sophisticated descriptions have been developed, incorporating Random Phase Approximation (RPA) calculations~\cite{nieves_inclusive_2011-2, PhysRevC.90.025501}, the plane-wave impulse approximation (PWIA) together with the use of realistic spectral functions~\cite{PhysRevC.100.045503} or relativistic distorted-wave impulse approximation (RDWIA) approaches~\cite{gonzalez-jimenez_nuclear_2019}.

%\sout{In the specific field of resonant production in the nucleon }{\ttblue (The DCC model describes either the lepton- and photon-induced single-pion production cross sections off the nucleon or inclusive lepton-and photon-induced inelastic cross sections off the nucleon. I am confused when you talk about 'resonant production'.)}\sout{, several groups have also developed sophisticated approaches to analyze and reproduce the nucleon structure in this regime,}
All previous nuclear models describing lepton-induced single-pion production (SPP) on the nucleus are based on the impulse approximation, i.e., they consider that the virtual boson couples only one nucleon in the nucleus. The description of the elementary boson-nucleon-pion vertex is done differently depending on the model considered.
%For the particular description of the inelastic inner structure of the nucleon, which needs to be included in the above-mentioned nuclear models, several groups have also developed different approaches. 
For example, the Hybrid model \cite{PhysRevD.76.033005,PhysRevD.95.113007,PhysRevD.97.013004} 
%\textcolor{red}{J: These are the reference related to Hybrid that are already in the article, same for EDRMF. Add others if you want to or is necessary} 
is used in the RDWIA approach~\cite{Nikolakopoulos_2023}, and the ANL-Osaka Dynamical Coupled-Channels model~\cite{nakamura_dynamical_2015,nakamura_impact_2019, DCConline,PhysRevC.67.065201,PhysRevC.88.035209} (DCC in what follows) has been recently implemented in the SuSAv2 framework~\cite{PhysRevD.108.113008}. 
The DCC has also been used in the extended factorization scheme~\cite{PhysRevC.100.045503}, which was incorporated in the ACHILLES event generator~\cite{Isaacson:2025cnk}; the Hybrid was implemented in the NuWro event generator using a local Fermi gas as nuclear model~\cite{Yan2024}.
A different approach for neutrino-induced single pion production is the MK model~\cite{kabirnezhad_single_2018}, which is used in the NEUT event generator~\cite{Hayato2021}.
% ~\cite{kabirnezhad_single_2018,kabirnezhad_single_2020,kabirnezhad_single_2023}
% {\ttblue I kept only reference to the MK model applied to neutrinos.} 

%{\ttgreen JA: I do not understand the previous sentence. Could you clarify?} {\ttorange The previous paragraph gives information about different nuclear models por CC1pi and this one does the same but at the nucleon level, i.e., it mentions some models to describe single-nucleon inelastic structure functions. We can rephrase it if it is not clear.}
%{\ttblue (It is unclear to me what is the difference between the models included in the previous paragraph and the two models of this paragraph.)}
%{\ttblue Previous sentence is also unclear to me, in particular: `inelastic inner structure of the nucleon, which needs to be included in...'. I suggest that when we talk about nuclear models (previous paragraph) we could say: All previous models nuclear models describing lepton-induced single-pion production on the nucleus are based on the impulse approximation, i.e., they consider that the virtual boson couples only one nucleon in the nucleus. Then, when we talk about the elementary vertex (this paragraph) we could say: The description of the elementary boson-nucleon-pion vertex is done differently by different models. Now we can talk about Hybrid, DCC and others. }{\ttorange (I have included this information above in the previous paragraph in orange. Check it and remove the coloured text and comments if you agree.)}

In a recent work \cite{PhysRevD.108.113008,PhysRevD.111.073002}, the superscaling model SuSAv2, developed for the charged-current quasielastic neutrino-nucleus cross section, was extended to the full inelastic regime (SuSAv2-inelastic model). In this framework, the DCC model
%, that describes the lepton-nucleon resonance channels, has been incorporated. The DCC model 
has been implemented~\cite{PhysRevD.108.113008} to give an accurate description of 
% resonant and nonresonant processes including interaction between channels and 
neutrino-induced one- and two-pion production, among other meson-production mechanisms in the resonance region. 
This model has been widely tested for photon, electron and neutrino scattering off single nucleons, providing information about the single-nucleon inelastic structure functions ($W_{1-5}$)~\cite{nakamura_dynamical_2015,nakamura_impact_2019, DCConline,PhysRevC.67.065201,PhysRevC.88.035209}. 
%\sout{Furthermore, this model not only gives the structure functions associated with the whole inclusive contribution (inclusive-DCC) that takes into account two-body and three-body meson-baryon final states (mainly $\pi N$, $\eta N$, $K\Lambda$, $K\Sigma$ and $\pi\pi N$)} \sout{ {\orange (check this and if anything is missing about definition of inclusive-DCC):} and the effects of the interference between the different, resonant and non-resonant, meson production channels, but also provides the functions connected to lepton-induced single pion production ($\pi$-DCC).}
%{\ttorange only from Delta??} {\ttorange In the plots we just call it $\pi$-DCC}.
This model accounts for the structure functions associated with the whole inclusive contribution and also the functions connected to lepton-induced SPP. In what follows, we refer to them as inclusive-DCC and $\pi$-DCC, respectively. The inclusive-DCC takes into account two- and three-body meson-baryon final states ($\pi N$, $\eta N$, $K\Lambda$, $K\Sigma$ and $\pi\pi N$) in a couple channel approach, while the $\pi$-DCC accounts only for a particular SPP channel. 
%{\ttorange (I agree with the new paragraph. We can remove the "sout" text.)}
Then, using $\pi$-DCC inelastic structure functions, we can distinguish between channels with a single $\pi^{+}$, $\pi^{0}$, or $\pi^{-}$ in the final state.   

%\sout{On the other hand, other models have tried to describe the neutrino-induced single-pion production (SPP) in the entire energy of interest for present and future accelerator based neutrino oscillation experiments in the context of RDWIA approaches.}
%the relativistic mean field (RMF) {\ttblue (I would prefer to talk about the RDWIA framework and then say that we use a RMF model for the bound state, but we can talk...)}. 
The Hybrid model, presented in~\cite{PhysRevD.95.113007}, aims at describing lepton-induced SPP in a broader energy region, so it could be used without phase-space restrictions in neutrino-oscillation experiments. 
It is based on the low-energy model of~\cite{PhysRevD.76.033005} and a high-energy Regge-based model, 
and can be used for SPP induced by electrons and charged- and neutral-current neutrinos. 
%\sout{The Hybrid model was later incorporated in the RDWIA nuclear framework, so it could be used to model cross sections for incoherent SPP on the nucleus \cite{PhysRevD.97.013004,Nikolakopoulos18,Nikolakopoulos_2023, PhysRevC.109.024608}. }
%\sout{It has been compared to MiniBooNE, MINERvA and T2K pion-detected cross section data. The tension observed in these studies between bubble-chamber and MINERvA data was shown in~\cite{Nikolakopoulos_2023}, using the RDWIA and also comparing with the relativistic plane-wave impulse-approximation (RPWIA).}
The Hybrid model was later incorporated in a relativistic mean-field nuclear framework, so it could be used to model cross sections for incoherent SPP on the nucleus \cite{PhysRevD.97.013004,Nikolakopoulos18,Nikolakopoulos_2023, PhysRevC.109.024608}.
It has been compared with MiniBooNE, MINERvA and T2K pion-detected cross section data using the relativistic plane-wave impulse-approximation (RPWIA)~\cite{PhysRevD.97.013004,Nikolakopoulos18} and RDWIA~\cite{Nikolakopoulos_2023}. The difference between RPWIA and RDWIA lies only in the description of the scattered nucleon, while the pion is treated as a plane wave in both cases.
Tension between bubble-chamber and MINERvA data was observed in~\cite{Nikolakopoulos_2023} with both approaches.

All these approaches are described in Sect. \ref{Formalism}, where the theoretical formalisms for the SuSAv2-inelastic and RDWIA frameworks are summarized. In Sect. \ref{Results}, we show the comparison between the predictions of the different theoretical models and the CC1$\pi$ experimental data provided by MiniBooNE (Sect. \ref{MiniBooNE}), MINERvA (Sect. \ref{MINERvA}) and T2K (Sect. \ref{T2K}), that operate at diverse kinematics and with different targets. In Sect. \ref{Conclusions}, we draw our conclusions.

\section{Formalism}\label{Formalism}

The superscaling approach (SuSA) is based on the scaling properties exhibited by inclusive electron scattering where the QE scattering cross section can be written, under certain conditions, as a term containing the single-nucleon cross section times a scaling function ($f$) that embodies the nuclear dynamics.
The analysis of inclusive electron scattering data~\cite{donnelly_superscaling_1999-2} has shown that for transferred momentum ($q$) values around 300 MeV/c or higher, 
%\sout{($q$ larger than about 400 MeV/c)}{\ttblue (most of the data of RL, used to extract $f_L^{exp}$ are for q lower than 400 MeV)}, 
the scaling function does not depend on $q$ (scaling of 1st kind) nor on the nuclear species (scaling of 2nd kind) and can therefore be expressed in terms of a single variable $\psi$, the so-called scaling variable.
 A more detailed description of superscaling can be found in \cite{donnelly_superscaling_1999-2,amaro_electron-_2020, amaro_neutrino-nucleus_2021, megias_inclusive_2016-1, megias_charged-current_2016, megias_neutrinooxygen_2018, PhysRevD.99.113002,megias_meson-exchange_2015}. 
 This approach has also been successfully applied to inclusive charged-current quasielastic (CCQE) neutrino scattering and, most recently, to the full inelastic regime for both electron and neutrino reactions. 
The corresponding model for the quasielastic region (SuSAv2-QE) is based on a set of QE scaling functions extracted from the relativistic mean field (RMF) theory~\cite{gonzalez-jimenez_extensions_2014}.

 The SuSAv2-inelastic model is an extension of the  SuSAv2-QE approach to the inelastic regime~\cite{gonzalez-rosa_susav2_2022,megias_charged-current_2017}. The double differential 
 %\sout{inclusive} {\ttorange [this inclusive (only lepton detected) is not the previous inclusive (all reaction channels). I would remove it]} 
 cross section for lepton-nucleus scattering with respect to the transferred energy $\omega$ and the scattered lepton solid angle $\Omega$ can be written in the general form~\cite{amaro_electron-_2020}
 \begin{equation}
\frac{d\sigma}{d\Omega d\omega} = \sigma_0 \sum_K v_K R^K \,,
 \end{equation}
 where $\sigma_0$ is an elementary cross section (the Mott cross section in the case of electron scattering), $v_K$ are lepton kinematic Rosenbluth factors and $R^K$ are the nuclear response functions, containing all the information about nuclear dynamics and the inner nucleon structure functions. The summed index $K$ is associated to different components of the nuclear tensor with respect to the direction of the transferred momentum ${\bf q}$.

%\sout{The nuclear responses depend on the transferred momentum and energy $(q,\omega)$, the exchanged four-momentum $Q^\mu$ and the invariant mass, $W_{X}$, of the hadronic final states or, equivalently, the dimensionless invariant mass $\mu_{X}\equiv W_X/m_N$. }
The inclusive nuclear responses, after integrating over the possible values of the invariant mass $W_X$ of the hadronic final states, depend only on the energy and momentum transferred to the nucleus.
%{\ttmag A.N.: (Inclusive responses only depend on $\omega,q$, the $W_X$ is integrated over below.)}

%{\ttorange (I have summarized the information and comments below as follows here. Check it and remove next comments in blue and orange in parenthesis after this text if you agree):} 

The invariant mass for lepton-nucleus reactions, $W_X$, can be determined via the 4-momentum of the final hadronic system $P_X$, which is defined by 4-momentum conservation as:
\begin{equation}
    k_i+P_A = k_f + P_{A-1} + P_X\,,
\end{equation} 
where $k_i$ and $k_f$ are the initial and final leptons 4-momenta, and $P_{A}$ and $P_{A-1}$ are the 4-momenta of the target and residual nucleus, respectively. Hence,
\begin{equation}
    W^2_X=P_X^2\,.
\end{equation} In the case of the SPP on nucleons, one has 
\begin{equation}
    W^2_X=(P_N+P_\pi)^2=(Q+P_i)^2=P_X^2\,,
\end{equation}
being $P_N$, $P_\pi$ and $P_i$ the 4-momentum of the final nucleon, produced pion and initial nucleon, respectively.
%$W_{X}\equiv m_N\mu_X$, of the hadronic final states. 
%{\ttorange I would just mention the invariant mass and after that define the dimensionless invariant mass.}

% {\ttblue (Can we define the invariant mass $W_X$? It is not trivial, not to me at least. In my opinion, it would not be necessary to define it if we were talking about scattering on a free nucleon, but this is scattering on a nucleus. To my understanding, the definition is as follows. 4-momentum conservation reads $$K_i+P_A = K_f + P_{A-1} + P_X,$$ where $K$ are lepton 4-vectors, $P_{A,A-1}$ the 4-vectors of the target and residual nucleus and $P_X$ the 4-vector of the nucleon that receives the hit. Hence, $$W^2_X=P_X^2.$$ In the case of SPP, one has $$W^2_X=(P_N+P_\pi)^2.$$ Is this correct? )}{\ttorange (It is not the definition that we use in the SuSAv2-inelastic or DCC codes as we do not have information about final-state hadrons. We can keep that equations but they are not valid for SuSAv2 codes so it could be better to just keep some references to that expressions and to the ones used for SuSAv2-inelastic.)} {\ttorange (SuSAv2 definition of the invariant mass included below in orange, so you can keep the other general definitions above these lines for the invariant mass and remove the comments here and below if you agree)}{\ttblue (I understand you don't use $W^2_X=(P_N+P_\pi)^2$ in SuSA but you do use the previous one, isn't it? In any case, my suggestion is we should define the variable that we use in SuSA.)} %\textcolor{red}{I do not understand giving a definition that is impossible to evaluate.}

In the SuSAv2-inelastic model \cite{gonzalez-rosa_susav2_2022,megias_charged-current_2017,PhysRevD.108.113008,PhysRevD.111.073002}, the nuclear responses are given by 
% \begin{widetext}
% \begin{equation}
\begin{align}
\label{RKinel}
&R^{inel}_{K}(q,\omega)=N\,\frac{2T_Fm_{N}^{3}}{k^{3}_{F}q}\,\non\\ 
&\times\int_{\mu_{X}^{min}}^{\mu_{X}^{max}}d\mu_{X}\mu_{X}f^{model}(\psi_{X})G^{inel}_{K} (q, \omega,W_{X})\,,
\end{align}
% \end{equation}
% \end{widetext}
being $\mu_{X}\equiv W_X/m_N$ the dimensionless invariant mass, $N$ the number of nucleons from the target involved in the reaction, $k_{F}$ the Fermi momentum and $T_{F}\equiv \sqrt{m_N^2 + k_{F}^{2}} - m_N$ the Fermi kinetic energy.
Thus, the inelastic nuclear responses are defined as the integral over all  possible final  hadronic states of the single-nucleon inelastic hadronic tensor $G^{inel}_{K}$ (characterized by the inelastic structure functions $W_{1-5}$) times the inelastic scaling function $f^{model}$ evaluated in a given nuclear model. The latter is written in terms of $\psi_X\equiv\psi_X(q,\omega,W_X)$, which is the extension of the QE scaling variable $\psi$ to the inelastic regime and now depends on the final state invariant mass $W_X$. The limits of the integral ($\mu_{X}^{min/max}$) can be adjusted depending on the kinematics  considered for a particular experimental or theoretical analysis %{\ttblue (the kinematics of the experimental sample one wants to compare to?)} \textcolor{red}{No. We use the experimental invariant mass for that.} 
and the limitations of the single-nucleon model used to characterize the hadronic tensor.
For example, in the case of the DCC model, there is a limitation up to an invariant mass of 2.1 GeV and $Q^{2}$ below 3 GeV$^{2}$~\cite{nakamura_dynamical_2015}. In general, the limits of this integral either in the SuSAv2 model or in a RFG framework can be defined as $\mu_X^{min}=1+m_\pi/m_N$ and $\mu_X^{max}=1 + \omega/m_N-E_{S}$, being $m_\pi$ the pion mass and $E_S$ the separation energy for a bound nucleon \cite{barbaro_2004}.
% {\ttblue (This is part of my problem. Here, when you talk about the invariant mass in the context of the DCC model it is clear that $W^2_X=(Q+P_i)^2=(P_X)^2$ where $P_{i,X}$ are the 4-momentum of initial nucleon and final hadronic system, respectively. In the SuSA context, the target is not a nucleon but a nucleus, that's why one needs to define what you call invariant mass in your model, otherwise one needs to guess it.)} {\ttorange (SuSAv2 definition of the invariant mass limits in the codes is included above in orange, so you can keep the other general definitions as well and remove the comments.)}
%{\ttorange I would add some general references in this paragraph. Some elements mentioned without reference or further details.}
%The limits of the integral ($\mu_{X}^{min/max}$) depend on the kinematics, on the 
%specific inelastic channel (full inelastic, DIS, RES, etc.) and on the range of validity of the inelastic structure functions used to evaluate the single-nucleon tensor. These limits can also be altered %to work with other models 
%to mix different models, avoiding double counting. 

\begin{figure*}[h]
\centering
  \includegraphics[width=1\textwidth]{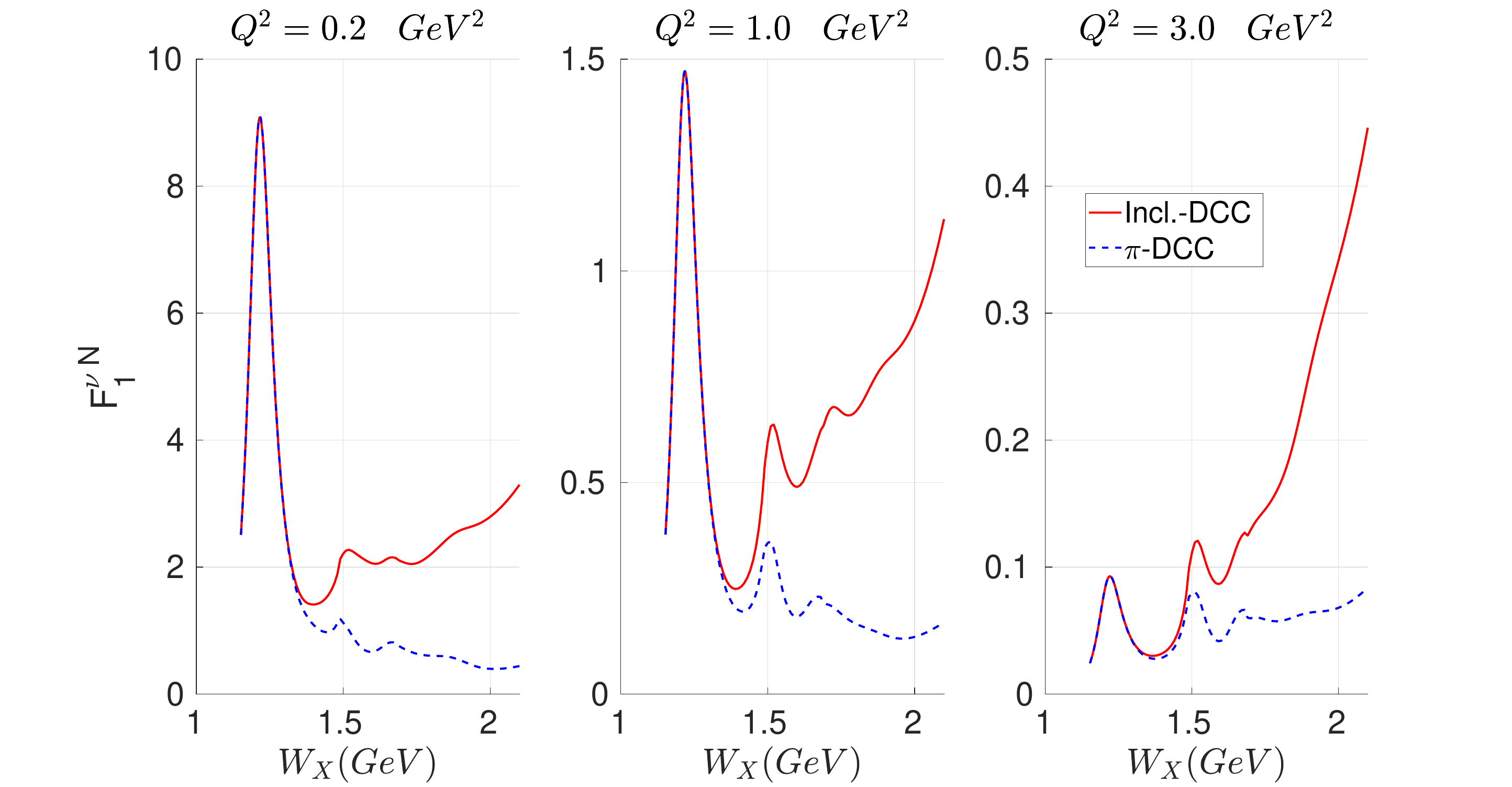}
    \caption{ Inclusive-DCC (red continuous line) and $\pi$-DCC (blue dashed line) inelastic structure functions $F_{1}=m_{N}W_{1}$ for neutrinos as a function of the invariant mass at different values of $Q^{2}$, namely 0.2 (left), 1 (center) and 3 (right) GeV$^{2}$.   
    }\label{FormFactor} 
\end{figure*}

In Fig. \ref{FormFactor}, we show the single-nucleon structure functions of the DCC model, displaying both $\pi$-DCC and inclusive-DCC contributions. 
%{\ttorange Have we mentioned that the structure functions go into the nuclear responses? Now yes.}. 
As expected, both curves match below the two-pion production threshold, in the region dominated by the $\Delta$ resonance, but as the invariant mass increases, the %\sout{only-pion} 
single-pion results ($\pi$-DCC) are below the inclusive ones that also take into account effects beyond SPP as mentioned above. 
%{\ttorange (I do not see the point of the next sentence that follows. It is not very clear to me and the definition of inclusive-DCC has already been given. We can reprhase it or remove it):} \textcolor{red}{This is due to the addition of reactions with double pions in the final state in the inclusive structure functions and to the effects of the interference between the different, resonant and non-resonant, pion production channels.}

%\textcolor{red}{Previous text}
%This is due to the other contributions beyond the production of pions from the $\Delta$ resonance {\ttorange only the Delta??} adding {\ttorange added?} to the structure functions; as such, this region is not composed only of the $\Delta$ resonance. {\ttorange maybe misleading or wrong. Check.}

Unlike the SuSAv2 model, in the RDWIA approach, the 
cross section describing the SPP process gives information not only about final-state leptons but also about final-state hadrons (pion and nucleon). It reads~\cite{gonzalez-jimenez_nuclear_2019}:
\begin{equation}
    \frac{d^{8} \sigma}{d E_{f} d \Omega_{f} dE_{\pi}d\Omega_{\pi}d\Omega_{N}}=\mathcal{F}\frac{k_{f}E_{f}p^{2}_{N}E_{\pi}k_{\pi}}{(2\pi)^{8}f_{rec}}l_{\mu \nu} h^{\mu \nu}\,,
\end{equation}
where
\begin{equation}
    f_{rec}=\frac{p_{N}}{E_{N}}\left( 1 + \frac{E_{N}}{E_{A-1}} \left| 1 + \frac{\textbf{p}_{N} \cdot (\textbf{k}_{\pi} - \textbf{q})}{p^{2}_{N}} \right| \right)\,.
\end{equation}

This expression is valid for SPP induced by electromagnetic as well as weak-neutral and weak-charged current interactions. It is written in terms of the lepton, outgoing pion and  outgoing nucleon variables. The factor $\mathcal{F}$ depends on the particular process under study and includes the boson propagator and the coupling constants at the leptonic vertex. The expression involves the leptonic tensor ($l_{\mu \nu}$) and the hadronic tensor ($h^{\mu \nu}$). 

% {\ttmag A.N. (I rewrote the following a bit, please check if you agree:)}
The hadronic tensor is defined as 
\begin{equation}
    h^{\mu \nu}= \sum_\kappa \sum_{m_{j}, s_{N}}[J^{\mu}(\kappa,m_j,s_N)]^*J^{\nu}(\kappa,m_j,s_N)\,.
\end{equation}
The outer sum runs over bound states labelled by $\kappa$. The total angular momentum of the bound state is $j = \lvert \kappa \rvert -1/2$ and the orbital angular momentum is $l = j\pm 1/2$ for $\kappa = \pm \lvert\kappa\rvert$. Here $m_{j}$ is the third component of the angular momentum, and $s_{N}$ is associated with the spin projection of the outgoing nucleon. 
We have suppressed here and in the following the explicit dependence on the four-momenta of the exchanged boson, the nucleon and the pion.
The tensor is written in terms of the 
%\sout{expected value} {
matrix elements of the hadronic current ($J^{\mu}$), which can be expressed as the Fourier transform of a current density
\begin{equation}
    J^{\mu}(\kappa,m_j,s_N) = \int \mathrm{d}^3\mathbf{r} e^{i\mathbf{q}\cdot\mathbf{r}} \mathcal{J}^\mu(\mathbf{r}, \kappa, m_j,s_N).
\end{equation}
In the relativistic impulse approximation it has the following structure
\begin{equation}
    \mathcal{J}^{\mu}(\mathbf{r})\sim \phi_\pi^{*}(\mathbf{r})\bar{\psi}_{N}(\mathbf{r},s_N)\hat{\mathcal{O}}^{\mu}_{1\pi}\psi(\mathbf{r}, \kappa, m_j)\,.
\end{equation}
Here $\psi_N$ and $\psi$ are, respectively, scattering and bound-state nucleon wavefunctions represented by four-component spinors, $\phi_\pi$ is the pion wavefunction, and $\hat{\mathcal{O}}_{1\pi}$ is a bilinear operator.
More details of the spinors and the operator can be found in \cite{PhysRevD.95.113007,PhysRevD.97.013004,Nikolakopoulos_2023}.
In particular, the calculations for charged pion production presented here are identical to those of~\cite{Nikolakopoulos_2023}, and $\pi^0$ production results were included in~\cite{Nikolakopoulos:2021nmb}.

Results in this work are obtained using the following approximations: the pion is treated as a plane wave, %$\hat{\mathcal{O}}_{1\pi}$ is the Dirac bilinear operator for free nucleons~\cite{PhysRevD.95.012010}, 
and the `local approximation' is invoked. The latter means that momenta that enter in the operator are fixed to their asymptotic values.
With these approximations the calculation of the current simplifies as explained in~\cite{Nikolakopoulos_2023}.
Recently, results without the local approximation and an analysis of the effects of invoking it were presented in~\cite{PhysRevC.109.024608}.
The outgoing nucleon wavefunctions are obtained with the energy-dependent RMF (EDRMF) potential from~\cite{gonzalez-jimenez_nuclear_2019}. 
In this approach, initial and final state potentials are identical for low nucleon energies, which leads naturally to Pauli blocking, and the conservation of the Dirac current. 
At higher nucleon kinetic energies the potentials soften, thereby avoiding unsound behavior caused by using an energy-independent potential~\cite{gonzalez-jimenez_constraints_2020}.

%{\ttorange This is also} combined with an energy dependence that softens nuclear potentials as the kinetic energy of the nucleon increases. 
%The Dirac spinors that describe the bound and scattered nucleons are denoted as $\psi$ and $\bar{\psi}_{N}(\equiv\psi^{\dagger}_{N}\gamma_{0})$, and $\phi$ is the wave function of the pion. $\mathcal{O}^{\mu}_{1\pi}$ is the SPP hadronic current operator. . 

% It provides an excellent description of data from small to large momentum transfers.

In the following section, we compare the predictions of the models described above for charged-current neutrino- and antineutrino-induced SPP off the nucleon and on the nucleus.

%from both SuSAv2-inelastic using $\pi$-DCC structure functions and RMF and contrast both models against the available CC1$\pi$ scattering data. 

\section{Results}\label{Results}

%\subsection{CCRES electron cross sections?}

%\textcolor{red}{I do not have specific results for electrons.}
%\subsection{CCRES neutrino cross sections}

In this section, we compare the predictions of the SuSAv2-DCC and EDRMF-Hybrid models with data on charged current neutrino-induced SPP. Our analysis focuses on channels that result in a single pion in the final state, where a significant contribution arises from the $\Delta$ resonance. The SuSAv2-inelastic model with $\pi$-DCC structure functions allows to separate the different SPP channels, i.e., $\pi^+$, $\pi^0$, or $\pi^-$. 
Similarly, the SPP channels are computed separately in the EDRMF-Hybrid model.

\begin{figure*}[h]
\centering
  \includegraphics[width=1\textwidth]{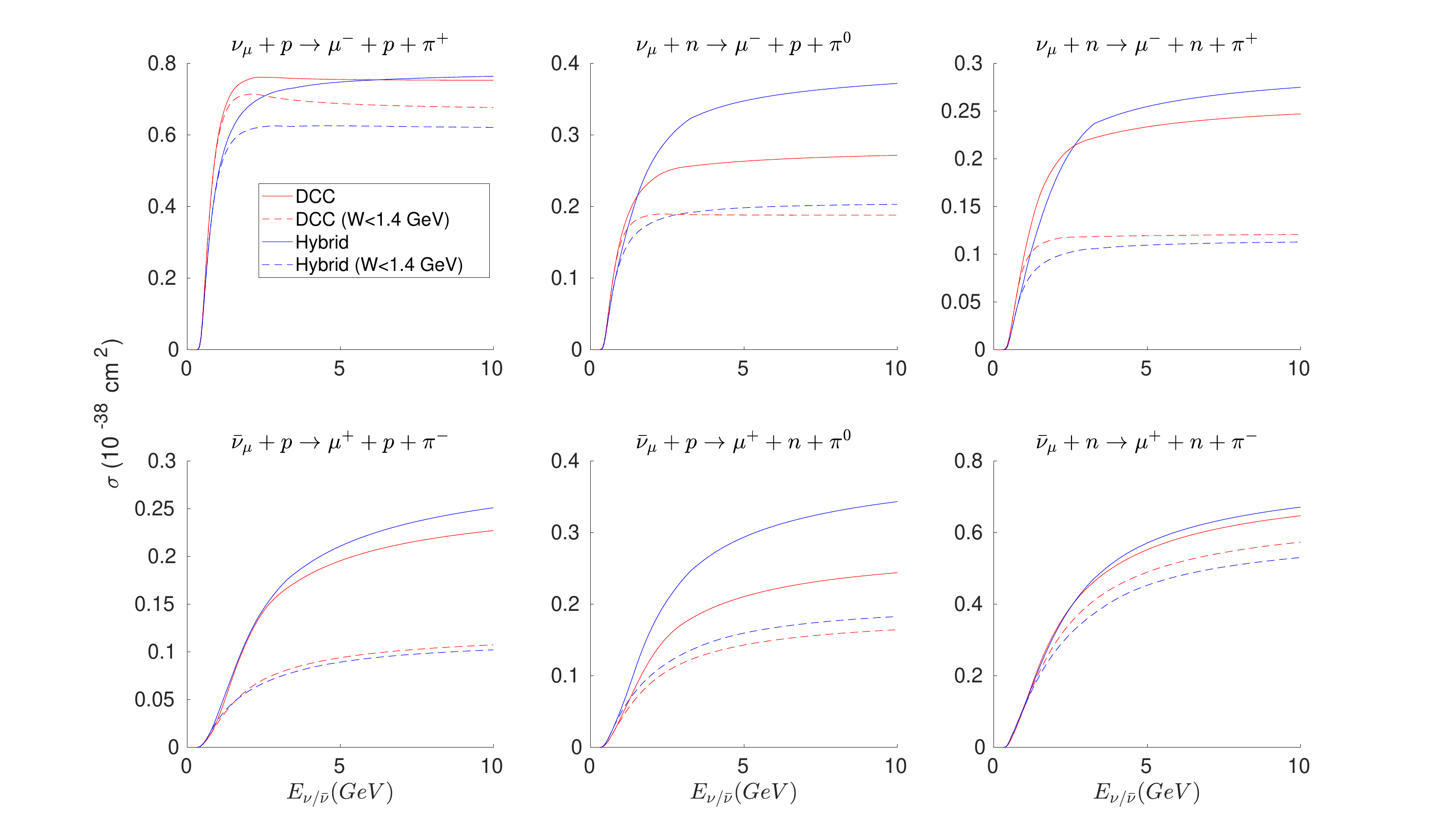}
    \caption{ CC1$\pi$ total cross sections ($10^{-38} cm^{2}$ ) in terms of the neutrino (top) or antineutrino (bottom) energies for different pion production channels: (top left) $p \rightarrow p + \pi^{+}$, (top center) $n \rightarrow p + \pi^{0}$, (top right) $n \rightarrow n + \pi^{+}$, (bottom left) $p \rightarrow p + \pi^{-}$, (bottom center) $p \rightarrow n + \pi^{0}$ and (bottom right) $n \rightarrow n + \pi^{-}$. Results from DCC and Hybrid models are shown, also considering a threshold of 1.4 GeV in the invariant mass.   
    }\label{totalxsec} 
\end{figure*}

%{\ttorange This next paragraph was in the previous section. Moved here.}
Before comparison to nuclear target data, we present in Fig. \ref{totalxsec}, the total cross sections for SPP off the nucleon for all neutrino and antineutrino channels obtained from the Hybrid and DCC models.
%arsented for all electroweak SPP channels. analyzed as a first step before their implementation within the SuSAv2 and EDRMF nuclear models, respectively, and their comparison with data in the following figures.  %{\ttorange we should briefly mention the motivation of this cut in W:} 
For $W_{X}<1.4$ GeV, where the main contribution comes from the $\Delta$ resonance, the prediction of the DCC model is higher than that of the Hybrid model, except for the $\pi^{0}$ channel. %\sout{in which case Hybrid is greater than DCC.} 
On the other hand, without this limitation in the invariant mass, the Hybrid model predicts larger total cross sections with the exception of the $\pi^{+}$ channel. %As such, in general, we will expect that the Hybrid model will provide a cross section larger than DCC in all channels except the $\pi^{+}$ production reactions.   

%{\ttorange Check this paragraph:} 
In Tables~\ref{multiprogram} and~\ref{multiprogram2}, the flux-folded CC1$\pi$ total cross sections for the different experiments are shown.
%(with the exception of MicroBooNE) are shown {\ttblue (why not?)}\textcolor{red}{There is no comparison with Hybrid in MicroBooNE. I will expect something similar to T2K due to the similitude of the fluxes.}. 
We consider both neutrino and antineutrino fluxes for MINERvA and the neutrino fluxes for T2K and MiniBooNE to analyze the flux-folded total cross sections for different neutrino- and antineutrino-induced SPP off free nucleons. 
%{\ttmag Of course, not all of these channels are relevant to the nuclear target data, e.g. neutrino cross sections averaged over the MINERvA antineutrino flux.} {\ttorange (tables updated for neutrinos and antineutrinos)}
This table shows the cross sections for $W_{X}<1.4$ GeV, where the $\Delta$ resonance dominates, and for $W_{X}<1.8$ GeV, which is a common threshold used in the MINER$\nu$A analyses. 
%\sout{In general, the results shown are consistent with those presented in Fig.~\ref{totalxsec} for the total cross section in terms of the neutrino energy, given that the MINER$\nu$A fluxes peak around 3.5 GeV, while the T2K and MiniBooNE ones peak below 1 GeV. } {\ttmag A.N. (Of course they are consistent ? they are calculated from this curve. Maybe you want to mention the mean energy difference in a different way ?)} {\ttorange (it has been included somehow in the next sentences. We can remove "sout" text and comments above if you agree)}
% 
In the case of MINER$\nu$A, we generally observe a notable difference between the total cross section from the DCC and Hybrid models, matching the differences shown in Fig.~\ref{totalxsec} at higher energies. 
For T2K and MiniBooNE, the models differ significantly for neutrino-induced $\pi^{+}$ production, which is also observed at energies around and below 1 GeV in the curves of Fig.~\ref{totalxsec} for that particular channel. 
% However, Hybrid and DCC 
The models tend to be more similar for the other pion production channels.
% , in accordance with Fig.~\ref{totalxsec}.    

% \begin{widetext}
  \begin{table*}
        \centering
        \begin{tabular}{c|c|c|c|c|c|c|}
            \cline{2-7}
             & \multicolumn{2}{|c|}{MINERvA $\nu$ flux}  
              & \multicolumn{2}{|c|}{T2K $\nu$ flux} 
               & \multicolumn{2}{|c|}{MiniBooNE $\nu$ flux} \\
            \cline{2-7}
             & $W_{X}<$1.4 GeV & $W_{X}<$1.8 GeV & $W_{X}<$1.4 GeV & $W_{X}<$1.8 GeV & $W_{X}<$1.4 GeV & $W_{X}<$1.8 GeV\\
            \hline
            \multicolumn{1}{|c|}{$\nu p \pi^{+}$(DCC)} &69.03 & 74.69 & 30.40 & 31.02 & 39.22  & 39.98 \\
            \hline
            \multicolumn{1}{|c|}{$\nu p \pi^{+}$(Hybrid)} & 62.12 & 72.03  & 24.9 &25.72 & 33.5 & 34.5   \\
            \hline
 \multicolumn{1}{|c|}{$\nu n \pi^{0}$(DCC)} &18.61 & 25.04  &7.95 & 8.64 & 10.18 & 11.01 \\
            \hline
 \multicolumn{1}{|c|}{$\nu n \pi^{0}$(Hybrid)} & 18.97 & 31.33 & 6.75 & 7.77 & 9.05 & 10.32\\
            \hline
\multicolumn{1}{|c|}{$\nu n \pi^{+}$(DCC)} &11.66 & 21.54  &4.52 &5.60 &  5.92 & 7.25 \\
            \hline
\multicolumn{1}{|c|}{$\nu n \pi^{+}$(Hybrid)} & 10.46 & 22.53 & 3.38 & 4.39 & 4.67 & 5.93 \\
            \hline            
        \end{tabular}
           \caption{Flux-folded CC1$\pi$ total cross sections ($10^{-40}$cm$^{2}$) for neutrino scattering on single nucleons and considering $W_{X}<$ 1.4 GeV and $W_{X}<$1.8 GeV, respectively. The kinematics of MINERvA, T2K and MiniBooNE considering their respective neutrino fluxes are analyzed comparing DCC and Hybrid results.\label{multiprogram}   %{\ttorange If I understand properly, you are using MINERvA, T2K and MiniBooNE neutrino fluxes for proceses with antineutrinos and viceversa, aren't you?}  \textcolor{red}{Yes. In the table, we use the neutrino flux for antineutrino processes and we use antineutrino flux for neutrino processes.}\label{multiprogram}  
           %{\ttorange More information is needed in the caption. Is this for single-nucleon for nucleus?..., etc. At some point (here or in the text) we should mention the motivation for the cuts in W at 1.4 and 1.8 GeV.}
           }
    \end{table*}
% \end{widetext}

% \begin{widetext}
  \begin{table}
        \centering
        \begin{tabular}{c|c|c|}
            \cline{2-3}
             & \multicolumn{2}{|c|}{MINERvA $\bar{\nu}$ flux} \\
             \cline{2-3}
            & $W_{X}<$1.4 GeV & $W_{X}<$1.8 GeV \\
            \hline            
\multicolumn{1}{|c|}{$\bar{\nu} p \pi^{-}$(DCC)}  & 7.76 & 15.72  \\ 
\hline
\multicolumn{1}{|c|}{$\bar{\nu} p \pi^{-}$(Hybrid)} &  7.46 & 16.69 \\ 
\hline
\multicolumn{1}{|c|}{$\bar{\nu} p \pi^{0}$(DCC)}  & 11.70 & 16.99 \\ 
\hline
\multicolumn{1}{|c|}{$\bar{\nu} p \pi^{0}$(Hybrid)}  & 13.16 & 23.44  \\ 
\hline
\multicolumn{1}{|c|}{$\bar{\nu} n \pi^{-}$(DCC)}  &  39.0 & 43.92 \\ 
\hline
\multicolumn{1}{|c|}{$\bar{\nu} n \pi^{-}$(Hybrid)} &  36.19 & 45.03   \\ 
\hline
        \end{tabular}
           \caption{Flux-folded CC1$\pi$ total cross sections ($10^{-40}$cm$^{2}$) for antineutrino scattering on single nucleons and considering $W_{X}<$ 1.4 GeV and $W_{X}<$1.8 GeV, respectively. The kinematics of MINERvA  considering the antineutrino flux are analyzed comparing DCC and Hybrid results.\label{multiprogram2}   %{\ttorange If I understand properly, you are using MINERvA, T2K and MiniBooNE neutrino fluxes for proceses with antineutrinos and viceversa, aren't you?}  \textcolor{red}{Yes. In the table, we use the neutrino flux for antineutrino processes and we use antineutrino flux for neutrino processes.}\label{multiprogram}  
           %{\ttorange More information is needed in the caption. Is this for single-nucleon for nucleus?..., etc. At some point (here or in the text) we should mention the motivation for the cuts in W at 1.4 and 1.8 GeV.}
           }
    \end{table}
% \end{widetext}

%{\ttorange I don't see the point of this paragraph. Somehow it is repeating previous information and the final sentence is more a conclusion that can be drawn after seeing the results. We could merge these sentences with the ones in the previous paragraph or remove them.} In the next subsections, we explore the limits of the models by comparing its predictions with data for different experiments, emphasizing those channels involved in the experiments that do not seem to be accounted for by the theoretical description. 
In the following subsections, we show the comparison between the SuSAv2 $\pi$-DCC and the EDRMF-Hybrid predictions and with CC1$\pi$ neutrino-nucleus data.
%{\ttorange Are all the kinematical cuts and restrictions applied to some of the following results mentioned in the paper? Check.}
%{\ttorange Is Table I cited in the text?}

\subsubsection{MiniBooNE} \label{MiniBooNE}

\begin{figure*}[!h]
\centering
  \includegraphics[width=\textwidth]{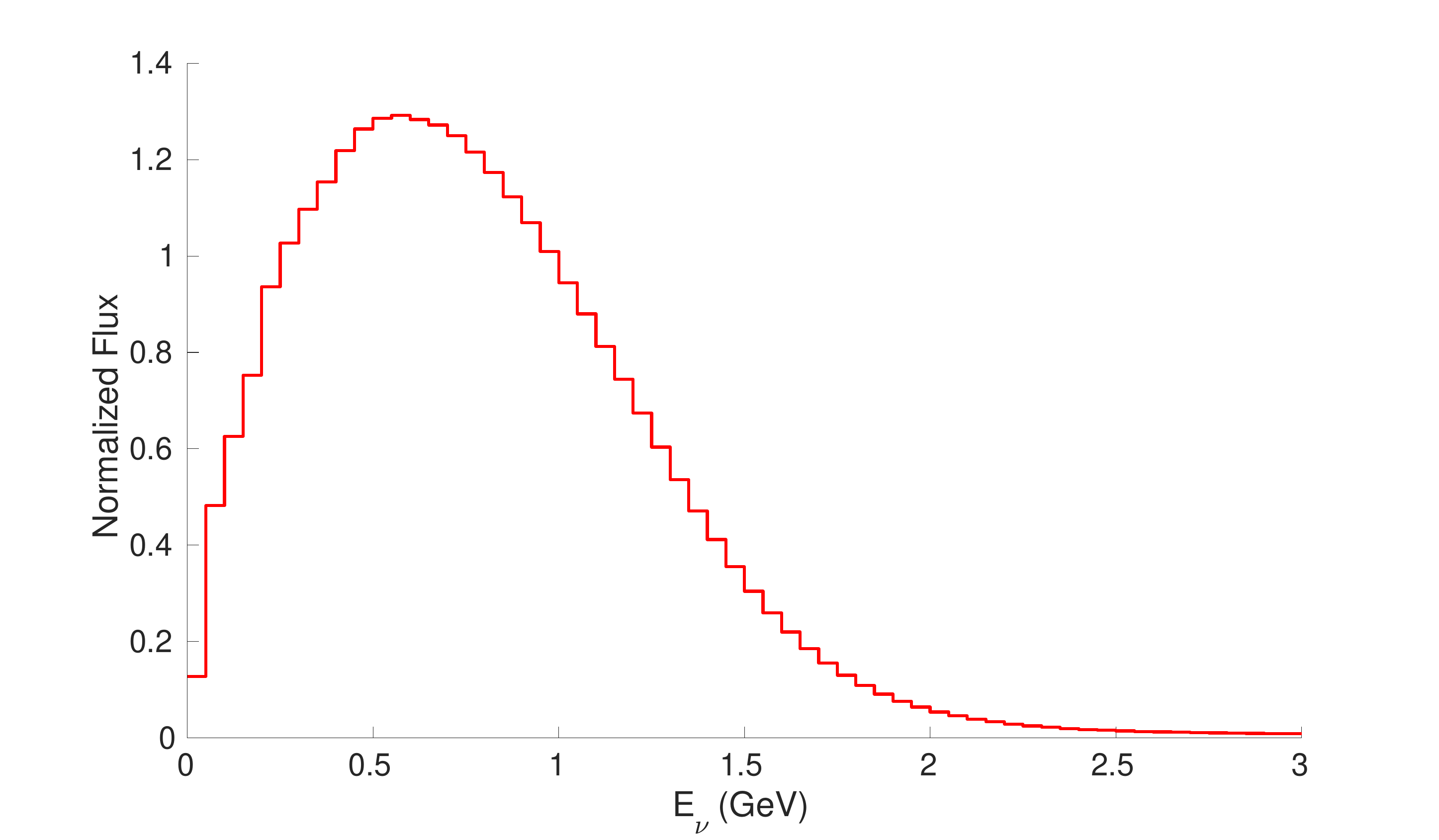}
    \caption{ Normalized neutrino flux from MiniBooNE~\cite{PhysRevD.83.052007,PhysRevD.83.052009} for neutrino-induced pion production processes.
    \label{Flux_MiniBooNE_Pion} }
\end{figure*}

The MiniBooNE experiment uses mineral oil, $\mathrm{CH}_{2}$, as a target. Fig.~\ref{Flux_MiniBooNE_Pion} shows the associated neutrino flux for CC1$\pi$ studies. 
%\sout{The profile  considered for pion production, limited in ranging from 0 to 3~GeV, although only values between 0.5 and 2~GeV are considered for the experimental analyses.}
The flux ranges from 0 to 3~GeV, although only values between 0.5 and 2~GeV are considered for the experimental analyses~\cite{PhysRevD.83.052007,PhysRevD.83.052009}.
The flux peaks around $E_\nu \sim 0.5~\mathrm{GeV}$. %\sout{For the complete flux, the average neutrino energy is approximately 1~GeV, and the flux drops to nearly zero at 2~GeV.}
%{\ttgreen I only see one profile} \textcolor{red}{Yes. At the end both profiles where the same. So, I just keep one.} {\ttorange (If I am not wrong, there were very minor differences in the plot between both fluxes that could be seen by eye. If you remove one curve, we have to rewrite this paragraph, but if both curves are exactly the same, it is fine as it is now, but we have to modify the text then.)} %{\ttorange I think the rest of the paragraph can be removed:} \textcolor{red}{In both cases, the only relevant part of the flux goes from 0.5 to 2 GeV.  he effects at low and high neutrino energy are not considered {\ttorange so you are not considering the low-energy tail (below 0.5 GeV) and high-energy tail (above 2 GeV) for pi+?? This is not clear in the text. Which cuts are you applying then?}} {\ttorange I do not think the previous sentence is correct. For pi+, the flux at 0.3 GeV, for example, is as relevant or more than the flux in the 1-2 GeV region.} {\ttorange The values mentioned in the plot 1GeV and 0.5-2.0 GeV, what do they mean? It is not clear and not stated in the figure or in the caption. I guess one is the mean energy and the other one is the range of Enu values. A bit misleading. This is mentioned here in the text but in the plot is not clear. I think we can remove it from the plot legend.} %In the case of $\pi^0$, the relevant neutrino energy range begins at 0.5~GeV and also decreases towards lower flux at higher energies.

\begin{figure*}[!h]
\centering
  \includegraphics[width=\textwidth]{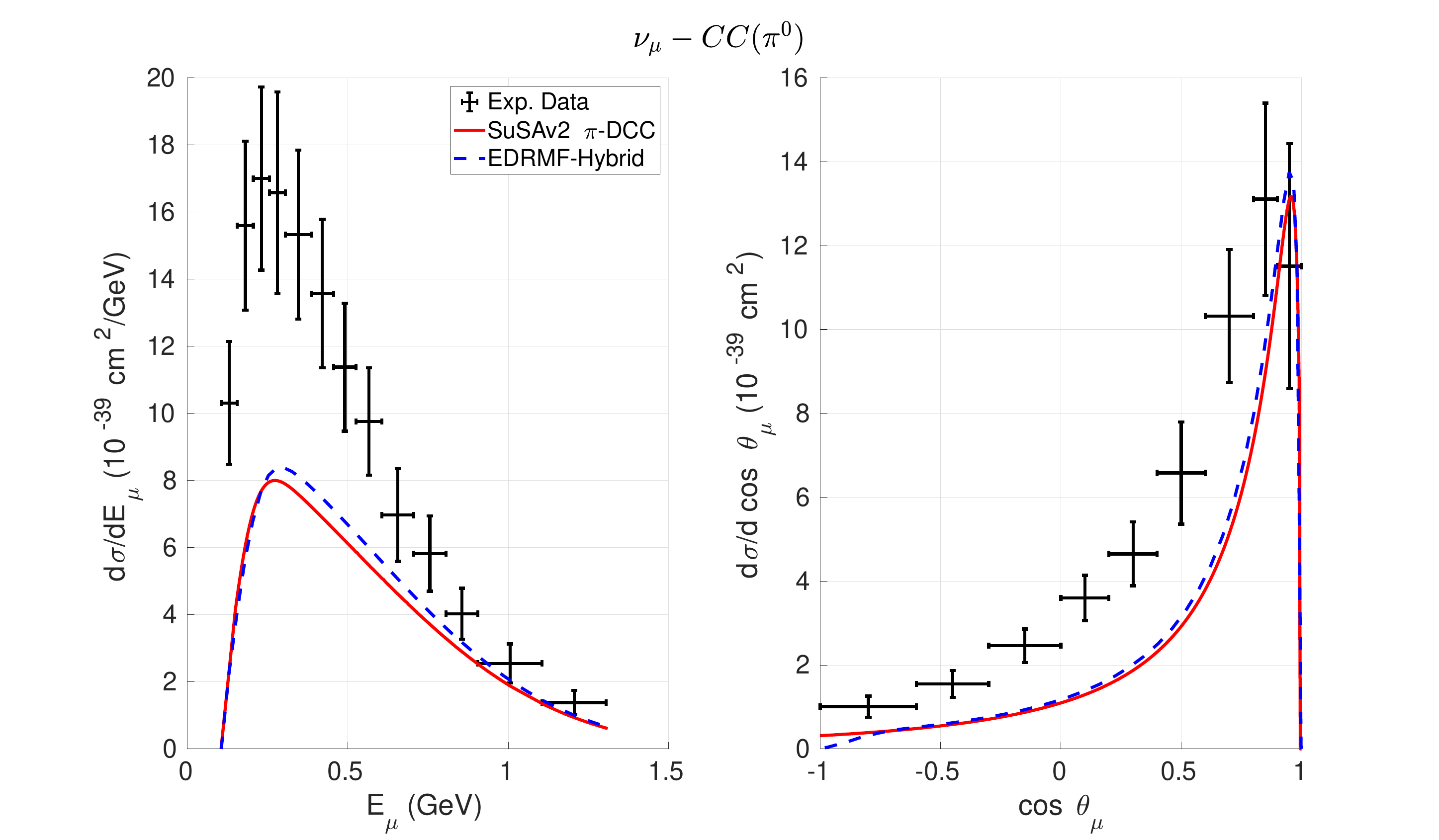}
    \caption{ MiniBooNE flux-averaged CC1$\pi^{0}$ $\nu_{\mu}$-CH$_{2}$ differential cross sections as a function of the muon energy (left) and the cosine of the scattering angle (right) compared with SuSAv2 $\pi$-DCC.  The flux is defined from 0.5 $< E_{\nu}/$GeV $<$ 2. The experimental MiniBooNE points are taken from~\cite{PhysRevD.83.052009}. 
    %{\ttorange No RMF results for MiniBooNE here?} \textcolor{red}{No RMF at the end. We can erase this Figure so we have comparison in all figure or argue lack of time in the text.}
    \label{MiniBooNE_PiZero} }
\end{figure*}

Fig.~\ref{MiniBooNE_PiZero} presents the CC1$\pi^0$ single-differential cross sections as a function of the muon momentum and the cosine of the muon scattering angle. In this channel, both the final-state muon and $\pi^0$ are detected. %\sout{ The total flux-averaged cross section is $0.25~\mathrm{cm}^2$, obtained by integrating either of the single-differential distributions.} 
The SuSAv2 $\pi$-DCC model in general underestimates the data.
% , it successfully reproduces the tail of $d\sigma/dp_\mu$ and the highest-angle bin in $d\sigma/d\cos\theta_\mu$.  
%In this approach, the tensor is evaluated using the $\pi$-DCC inelastic structure functions which model pion production with one proton and one 
%\sout{pion zero}
%$\pi^0$ in the final state. {\ttblue (Can we remove last sentence? The model was already explained somewhere else.)} {\ttorange I agree. Removed.}
%\sout{The model prediction does not include coherent pion production, though according to ~\cite{PhysRevC.90.025501} its contribution is negligible in this case.}
%its contribution is negligible in these kinematics~\cite{PhysRevC.90.025501}. 
%{\ttblue (What do you mean by "is excluded"? And we need a reference to say that coh. SPP is small. So, maybe something like: The model prediction does not include coherent pion production, though according to Ref.~\cite{} its contribution is negligible in this case.)} {\ttorange (I agree with the proposed sentence. Once we have the reference, we can update the text and remove comments.)} {\ttgreen O.K. Just add the reference in the above text}
In a similar way,  %\sout{pion-exchange} 
intra-nuclear cascade effects (e.g., a $\pi^+$ converting into a $ \pi^0$), are not included in the model which could explain part of the observed underestimation as discussed in~\cite{PhysRevD.97.013004,PhysRevD.110.092014}. According to \cite{PhysRevD.97.013004}, the MiniBooNE cross section would not be significantly modified; on the contrary, the MicroBooNE analyses of \cite{PhysRevD.110.092014} show an increase of around 50\% due to these effects. Thus, further studies concerning intra-nuclear cascade effects are required before further conclusions can be drawn.
%\textcolor{orange}{(G: Do we need this final sentence? Somehow it is repeating the previous text.)} \textcolor{red}{(I think is necessary. We present the other studies in the previous sentence and we draw our conclusions in this one.) \textcolor{orange}{(Ok... I have added a sentence above to remove this next one. Check it.)}} \sout{As such, we do not have a concrete prediction and we cannot completely disregard the possible effects that the intra-nuclear cascade effects have in the final result.}} 
%{\ttblue (it would be nice to add references here to support this statement)}. 
Another reason for this discrepancy can be related to the differences between DCC and Hybrid,
producing the latter larger results for CC1$\pi^0$, unlike other channels, as observed in Fig.~\ref{totalxsec}.
However, at the relatively low energies of MiniBooNE, the differences between DCC and Hybrid are not expected to be as significant as for other higher-energy experiments, such as MINERvA, as will be discussed later. %{\ttblue (it is unclear which effect you mean here, the pi+ going to pi0 or the difference between DDC and hybrid?)}. 
%{\ttorange (Text modified, but it should be reviewed once we add the EDRMF curve to this plot.)}. 
%{\ttv Is the term "pion-exchange cascade" normally used for this process? Usually pion-exchange means a pion exchanged between two nucleons. This process is rather a final state interaction which changes the charge of the pion.} {\ttorange (it has been rephrased).} 
In general, the contribution given by SuSAv2 $\pi$-DCC accounts for roughly half of the total observed CC1$\pi^0$ MiniBooNE result.

%\textcolor{red}{Previous paragraph. It has been rewritten}
%This prediction includes only the contribution of the $\Delta$ resonance {\ttorange (not true, already discussed. This paragraph will be rewritten)}, where a neutrino interacts with a neutron to form a $\Delta^+$ that subsequently decays into a proton and a $\pi^0$. Other mechanisms, such as nonresonant pion production, contributions from resonances other than the $\Delta$ one, or pion-exchange cascade effects (e.g., a $\pi^+$ converting into a $ \pi^0$), are not included in the model. Coherent pion production is also excluded, though its contribution is negligible in these kinematics. In general, $\Delta$-mediated pion production accounts for roughly half of the total observed $\pi^0$ yield.

The following figures (Figs.~\ref{MiniBooNE_PiPlus_SD} to~\ref{MiniBooNE_PiPlus_DD_2}) show results for CC1$\pi^+$ cross sections as a function of the muon kinetic energy, using the same flux shown in Fig.~\ref{Flux_MiniBooNE_Pion}. In this case there is an experimental cut in the phase space, so only processes with $W^{exp}<1.35$ GeV %\textcolor{orange}{why 1.35 here and 1.4 in the tables?} \textcolor{red}{J: This is  the cut given by the MiniBooNE article. The cuts in the table are about the invariant mass not the experimental invariant mass and there were shown to contrast DCC and Hybrid.} 
are considered, being $W^{exp}\equiv\sqrt{m_{N}^{2} +2m_{N}\omega -|Q^{2}|}$. 
% where the contribution of the $\Delta$ resonance is dominant.
In Fig.~\ref{MiniBooNE_PiPlus_SD}, the single-differential cross section is presented. Here, various pion production channels contribute, namely $\pi^+$ production from neutrino interactions with both neutrons and protons from carbon.
% , which are addressed within SuSAv2 $\pi$-DCC. The 
%In which, we consider the channels from the DCC which consider $\pi^{+}$ with protons or neutrons in the final state. 
%The SuSAv2 $\pi$-DCC also includes the 
The contribution of protons from the hydrogen component ($\mathrm{H}_2$) of the target is also considered.
% by evaluating the proton contribution of the structure functions. 
The DCC model accounts for most of the cross section strength, but contributions from coherent and %\sout{cascade} 
 pion rescattering processes are missing. %\textcolor{orange}{(G: Modified. Check it.) Comment: 15 percent is for coherent or pion rescattering? because later you mention 10 percent for the coherent pion production which is not the same. It is not clear what do you mean in each case.} 
 Pion rescattering may reduce the cross section by approximately 15\% \cite{PhysRevD.97.013004}. In this work, coherent pion production is not included; it is expected to contribute at most about $10\%$ of the cross section~\cite{PhysRevC.90.025501}. %,PhysRevLett.113.261802}  % could explain why the SuSAv2 $\pi$-DCC model slightly underpredicts the data {\ttorange where is the underprediction in Fig. 5??}.
On the other hand, the EDRMF-Hybrid results are similar at these kinematics to the SuSAv2-DCC ones, being the latter approximately 5\% higher, which is consistent with the results shown in Fig.~\ref{totalxsec}. 
%{\ttorange meaning of the next sentence?:} It is not possible to discriminate between both models at these particular kinematics and observable. }

%\textcolor{red}{Previous Text}
%{\ttorange (all the discussion mentioning DCC as only Delta is wrong)}
%In Fig.~\ref{MiniBooNE_PiPlus_SD}, the single-differential cross section is presented. Here, various $\Delta$-driven channels are considered, including $\pi^+$ production from neutrons and protons in the carbon nucleus, which are addressed within SuSAv2 $\pi$-DCC, as well as contributions of protons from the hydrogen component ($ \mathrm{H}_2$) of the target, which are included considering only the proton contribution of the structure functions. The $\Delta$ resonance accounts for most of the cross section strength, but contributions from coherent, nonresonant, other resonant, and cascade processes are missing. This could explain why the SuSAv2 $\pi$-DCC model slightly underpredicts the data.

\begin{figure*}[!h]
\centering
  \includegraphics[width=\textwidth]{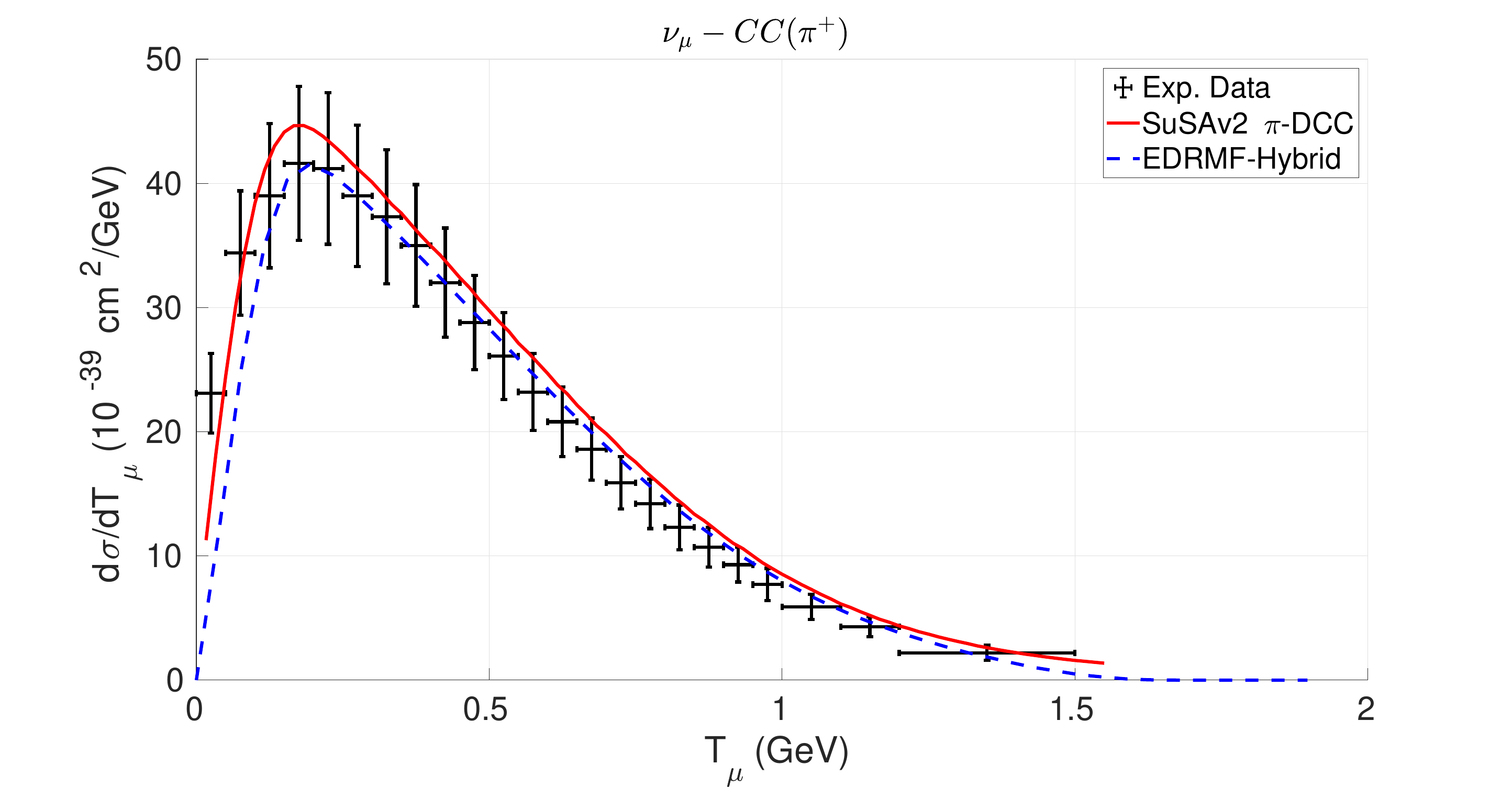}
    \caption{ MiniBooNE flux-averaged CC1$\pi^{+}$ $\nu_{\mu}$-CH$_{2}$ differential cross section as a function of the muon kinetic energy compared with SuSAv2 $\pi$-DCC and EDRMF-Hybrid. The flux is defined from 0.5 $<E_{\nu}/$GeV $<$ 2 and $W^{exp}<1.35$ GeV. The MiniBooNE data are from~\cite{PhysRevD.83.052007}.\label{MiniBooNE_PiPlus_SD}}%{\ttorange I would try to keep some consistency in the plots when possible. Here and in other plots, SuSAv2 and DCC are red solid lines while EDRMF and Hybrid curves are dashed blue, but this change in other plots.}
\end{figure*}

Fig.~\ref{MiniBooNE_PiPlus_0p95to1} shows the double-differential cross section in the very forward angular region. This region is more sensitive to nuclear ground state effects, like Pauli blocking and Fermi motion. At low muon kinetic energies, the absence of the additional coherent and
%\sout{cascade} 
pion rescattering effects 
%{\ttblue (I've changed a few times `cascade effects' by pion rescattering effects, please check if you agree. Cascade is a way of modeling some effects, but the effects are something else.)} 
could lead to a more pronounced underestimation of the data ~\cite{PhysRevC.90.025501}.
%{\ttblue (reference(s) in which coherent contribution is shown)}. 
These discrepancies between theory and data diminish as the muon kinetic energy increases, observing a good agreement with data for the EDRMF model at large $T_\mu$ values and a slight overestimation when compared with SuSAv2. 
%This can be related not only to the DCC vs Hybrid differences but also to treatment of the very low energy and momentum transfer, which is more accurate in the EDRMF. {\ttblue (I changed a little bit the last sentence, trying to keep the original message, but I would simply remove it completely. The two models are too close to each other, and high $T_\mu$ implies low q and omega? It could be same q,omega but larger neutrino energies...)}
% the scaling violation effects in the SuSAv2 approach at very low momentum and energy transfer, 
 %\textcolor{red}{Both models have the same tendencies, and we cannot substantially differentiate between them} {\ttorange but we can observe some differences between them... (I do not understand the previous sentence then)} .

\begin{figure*}[!h]
\centering
  \includegraphics[width=\textwidth]{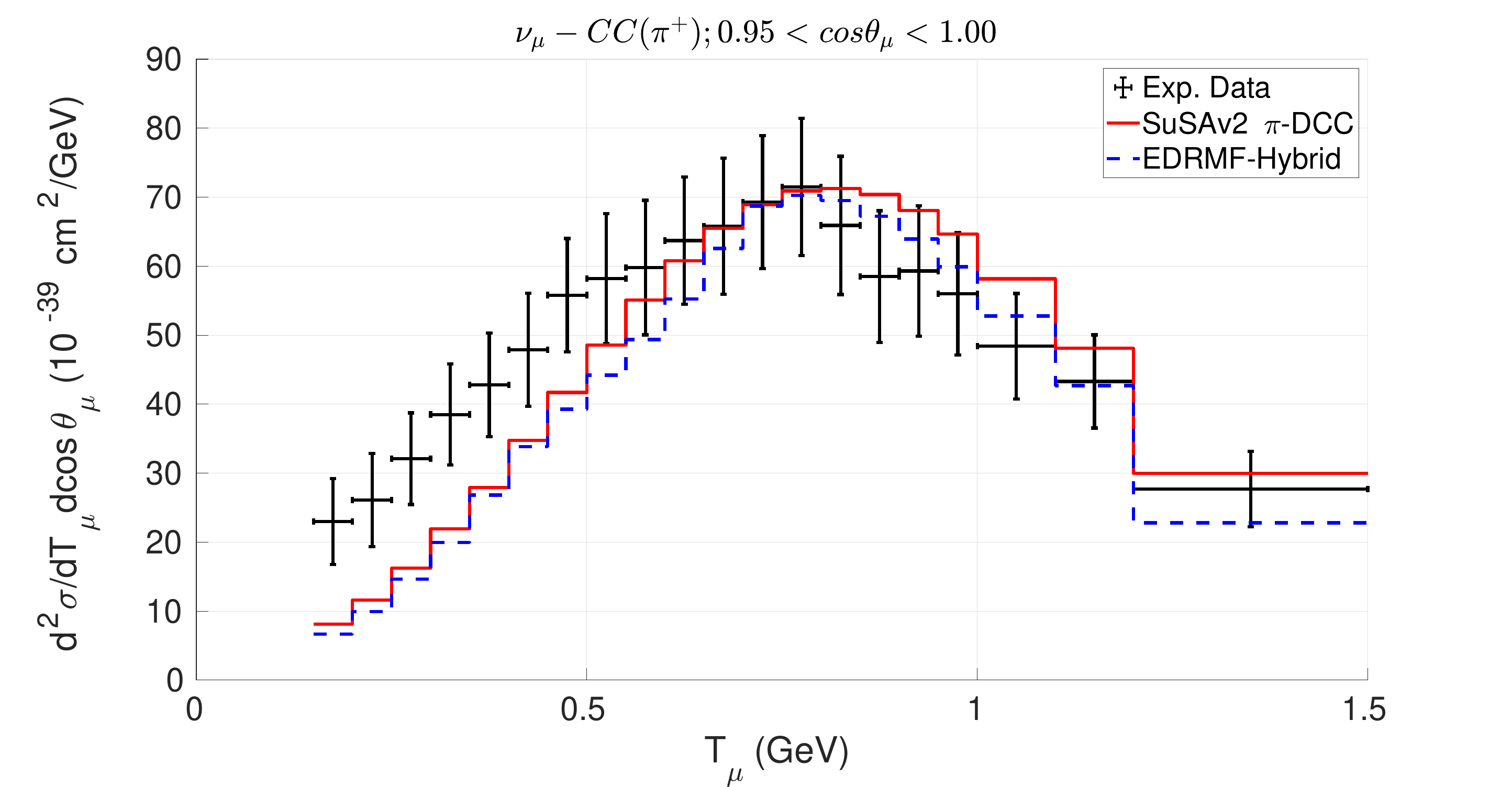}
    \caption{ MiniBooNE flux-averaged CC1$\pi^{+}$ $\nu_{\mu}$-CH$_{2}$ double-differential cross sections for  $0.95<\cos\theta < 1$ as a function of the muon kinetic energy compared with SuSAv2 $\pi$-DCC and EDRMF-Hybrid. The flux is defined from 0.5 $<E_{\nu}/$GeV $<$ 2 and $W^{exp}<1.35$ GeV. The experimental MiniBooNE points are taken from~\cite{PhysRevD.83.052007}. \label{MiniBooNE_PiPlus_0p95to1}}
\end{figure*}

In Figs.~\ref{MiniBooNE_PiPlus_DD} and~\ref{MiniBooNE_PiPlus_DD_2}, we present the charged-current single $\pi^+$ (CC1$\pi^+$) double-differential cross section as a function of the cosine of the muon scattering angle, averaged over bins of muon kinetic energy. 
% These results are consistent with and complement our findings in previous figures. 
In general, we observe good agreement with data for both models except at very forward angles and low-intermediate $T_\mu$ where the models underestimate the data; at large $T_\mu$ the data are overestimated. The agreement with data is slightly better for the EDRMF-Hybrid.
%{\ttorange I do not see this sentence when checking the plots:} However, the SuSAv2 $\pi$-DCC and RMF model generally underestimates the data {\ttorange where? is that true? I actually see some slight overestimations and some underestimations in other cases.}, particularly at very forward angles. This result is in accordance {\ttorange (it has nothing to do actually)} with the behavior observed in Fig.~\ref{MiniBooNE_PiZero}.

\begin{figure*}[!h]
\centering
  \includegraphics[width=\textwidth]{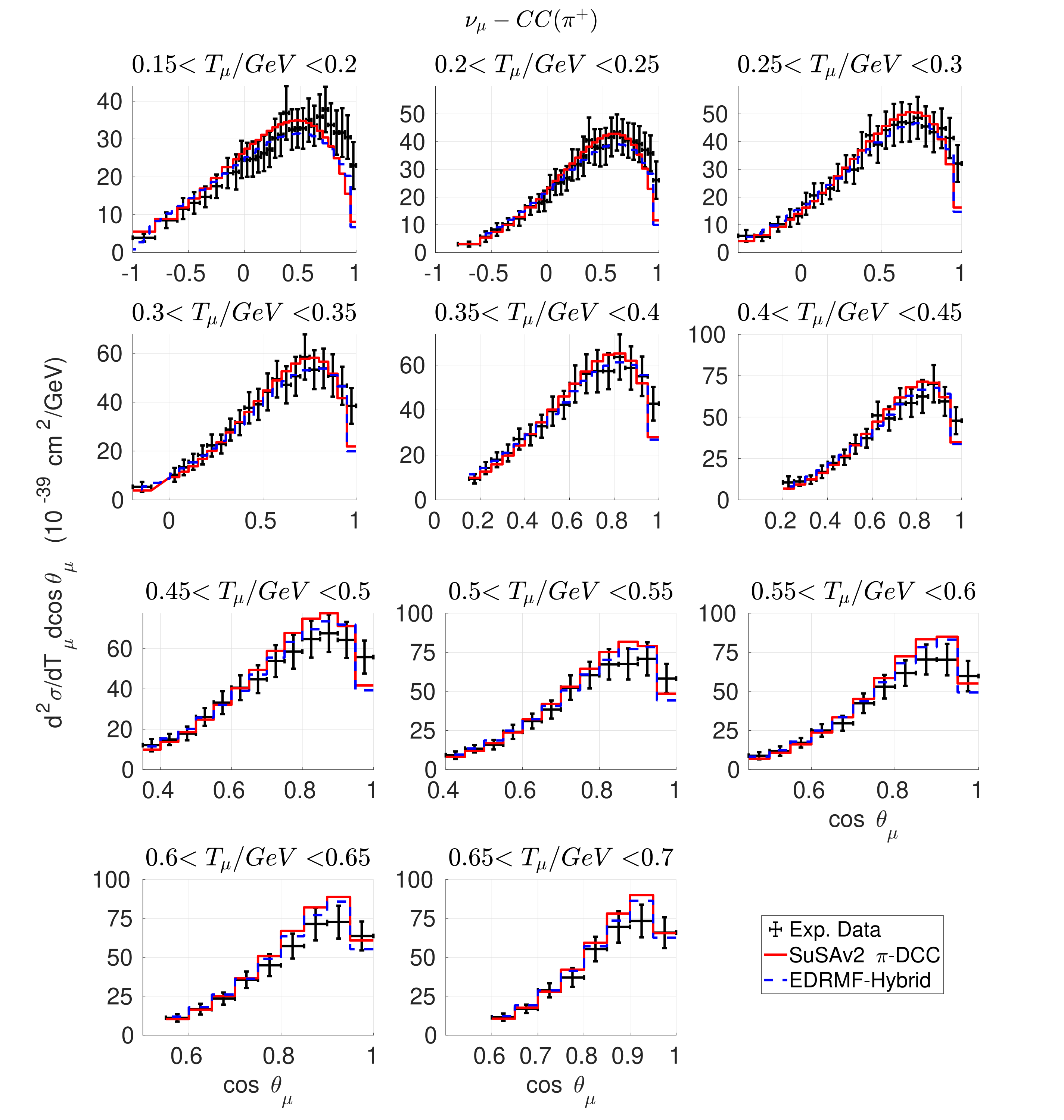}
    \caption{ MiniBooNE flux-averaged CC1$\pi^{+}$ $\nu_{\mu}$-CH$_{2}$ double-differential cross sections for several values of muon kinetic energy as a function of the muon scattering angle. Data are compared with the SuSAv2 $\pi$-DCC and EDRMF-Hybrid. 
    The flux is defined from 0.5 $<E_{\nu}/$GeV $<$ 2 and $W^{exp}<1.35$ GeV.  The MiniBooNE data are from~\cite{PhysRevD.83.052007}. \label{MiniBooNE_PiPlus_DD} }
\end{figure*}

\begin{figure*}[!h]
\centering
 \includegraphics[width=\textwidth]{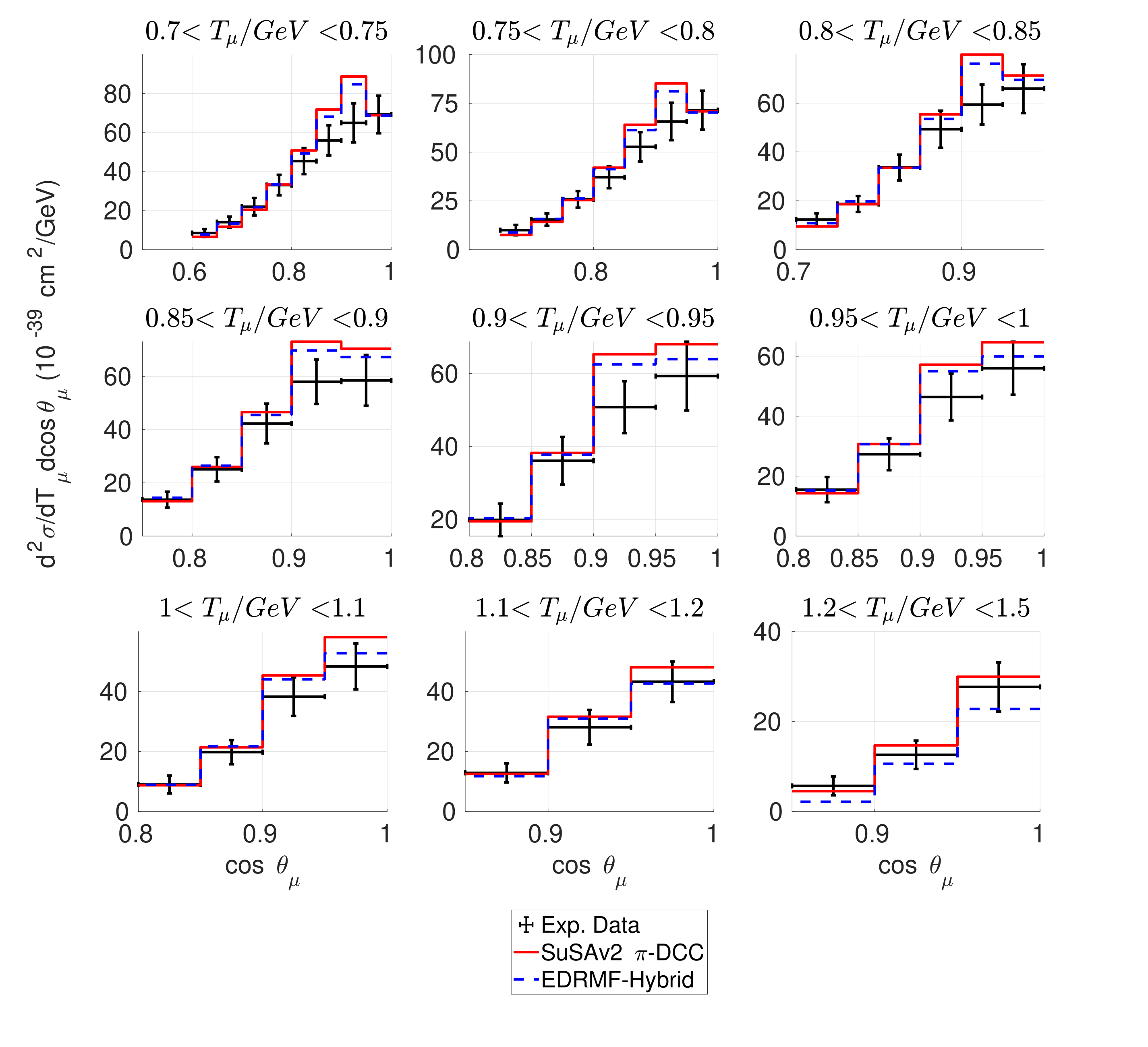}
    \caption{ MiniBooNE flux-averaged CC1$\pi^{+}$ $\nu_{\mu}$-CH$_{2}$ double-differential cross sections for several values of muon kinetic energy as a function of the muon scattering angle  compared with SuSAv2 $\pi$-DCC and EDRMF-Hybrid. The flux is defined from  0.5 $<E_{\nu}/$GeV $<$ 2 and $W^{exp}<1.35$ GeV.  The experimental MiniBooNE points are taken from~\cite{PhysRevD.83.052007}. \label{MiniBooNE_PiPlus_DD_2}}
\end{figure*}

\subsubsection{MINERvA} \label{MINERvA}

\begin{figure*}[!h]
\centering
  \includegraphics[width=0.9\textwidth]{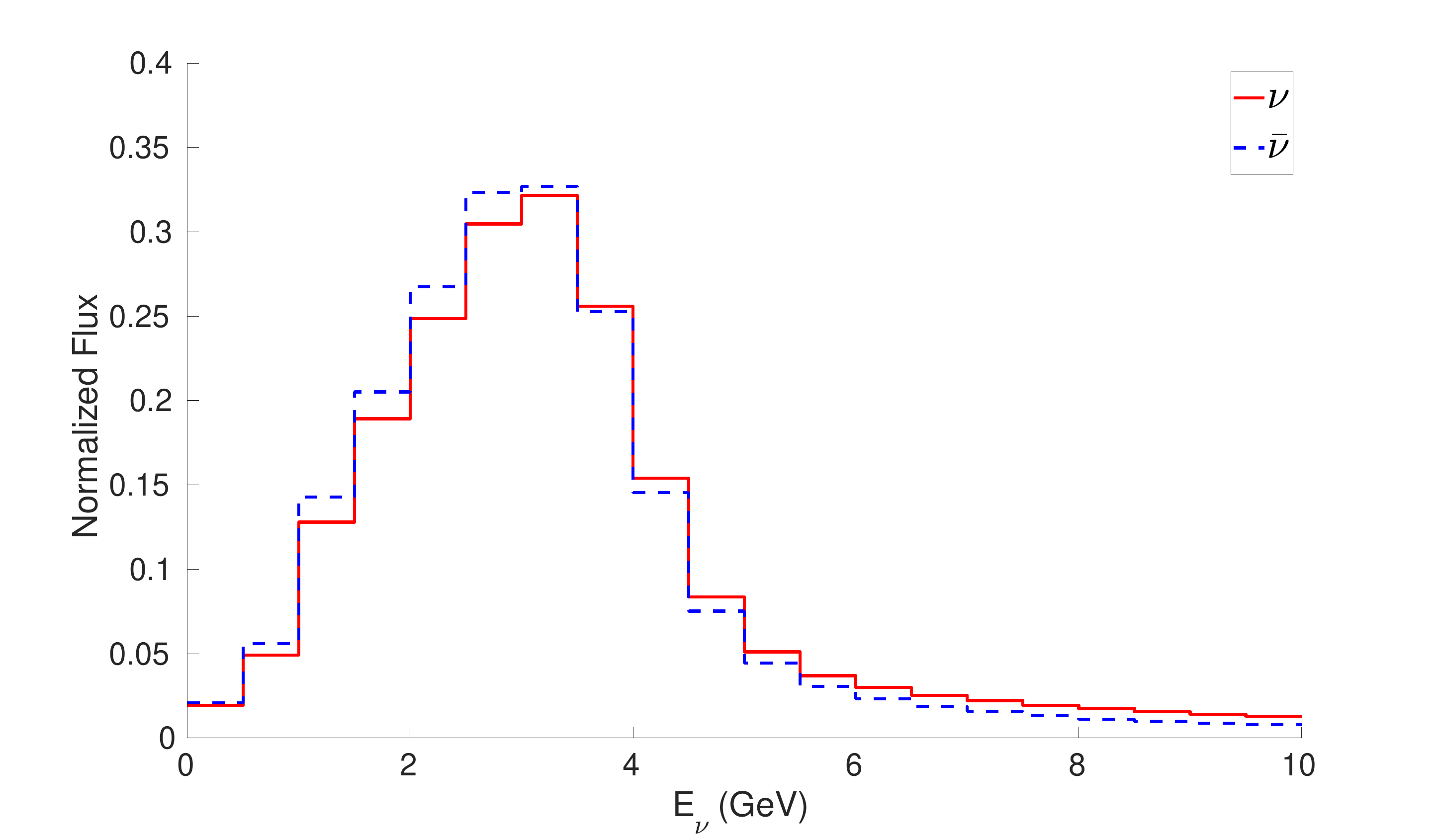}
    \caption{ Normalized neutrino (red line) and antineutrino (blue dashed line) fluxes from MINERvA~\cite{PhysRevD.94.052005,PhysRevD.96.072003,PhysRevD.100.052008} for neutrino-induced pion production processes. \label{Flux_MINERvA_Pion}}
\end{figure*}

In this experiment, the target material is hydrocarbon ($\mathrm{CH}$). As shown in Fig.~\ref{Flux_MINERvA_Pion}, the flux for the different pion production channels peaks at approximately 3.5~GeV and extends up to 20~GeV, although it becomes negligible above 7~GeV. This energy range is significantly larger than that of the MiniBooNE experiment discussed previously.

\begin{figure*}[!h]
\centering
  \includegraphics[width=0.9\textwidth]{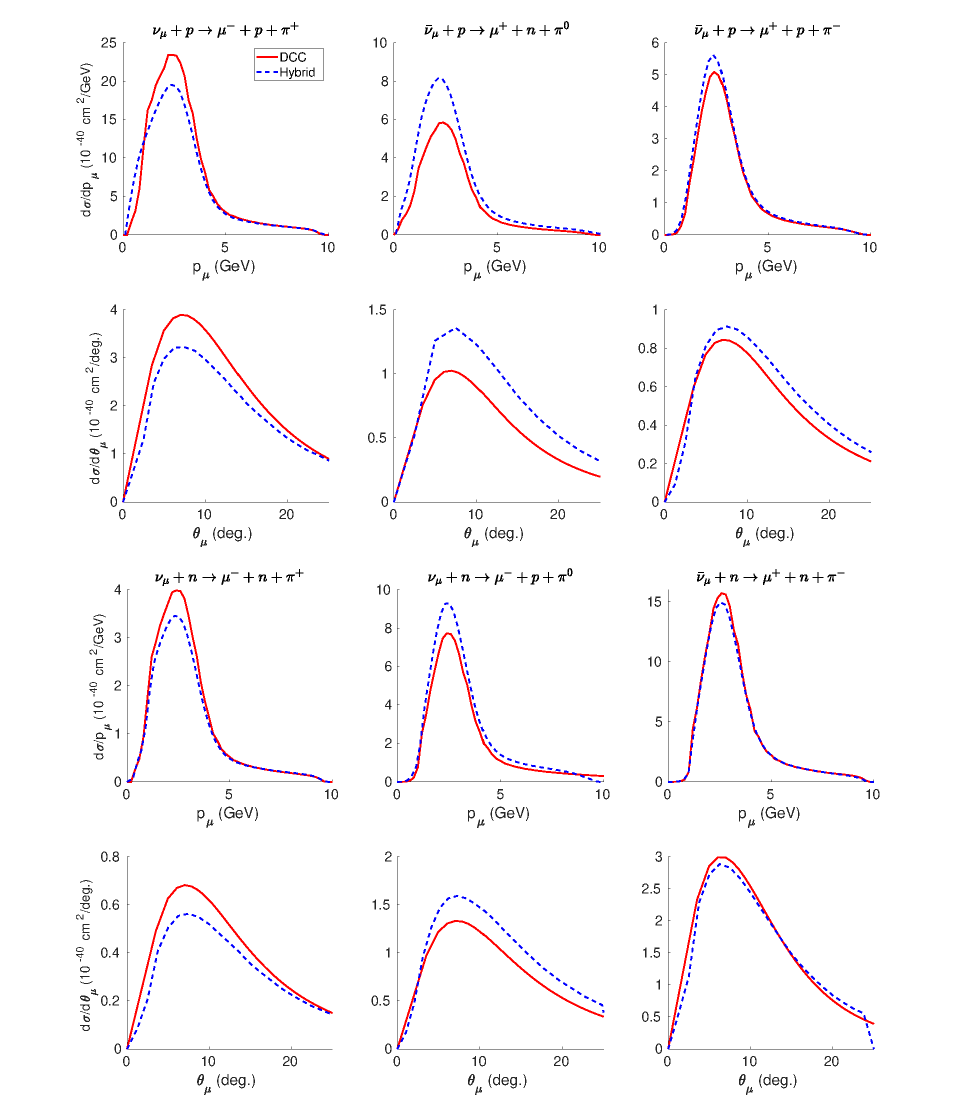}
     \caption{  MINER$\nu$A flux-folded CC1$\pi$ single-differential cross sections in terms of the muon momentum (top panels) and scattering angle (bottom panels) for different neutrino- and antineutrino-induced pion production scattering off single nucleons.  (Left) $\pi^+$ processes for 1.5 $<E_{\nu}/$GeV $<$ 10 and $W^{exp}<$ 1.4 GeV; (Center Top) $\bar{\nu}\pi^0$ processes for 1.5 $<E_{\nu}/$GeV $<$ 10 and $W^{exp}<$ 1.8 GeV; (Center Bottom) $\nu\pi^0$ processes for 1.5 $<E_{\nu}/$GeV $<$ 20, $\theta_{\mu}<$ 25 deg.~and $W^{exp}<$ 1.8 GeV; and (Right) $\pi^-$ processes for 1.5 $<E_{\nu}/$GeV $<$ 10, $\theta_{\mu}<$ 25 deg.~and $W^{exp}<$ 1.8 GeV. %{\ttmag A.N. (I sent you also the other channels: note my comment in the email: the cross sections in this figure are divided by a factor 13. For example, from the $\pi^+$ curve one estimates the total cross section (as a triangle) $\sigma \approx 5*2/2 = 5 $. The actual total cross section, see the table, is larger $\sigma \sim 5*13 = 65 $. I suggest removing the factor $1/13$ and the label 'per nucleon' or explaining this factor in the caption clearly.) {\ttorange (The restriction of 25 deg. is only for some panels or for all of then? If for all of them, it should be updated here and in the next plots.)}} \textcolor{red}{The restriction of 25 deg. only affects neutrino pion zero and antineutrino pion minus.}
     \label{comparison_proton_minerva}}
\end{figure*}

%\begin{table}
 %   \centering
%\begin{tabular}{ |c|c|c|c|c| } 
% \hline
%   & $\nu p \rightarrow p\pi^{+}$ & $\bar{\nu}p \rightarrow n\pi^{0}$ & $\nu n \rightarrow p \pi^{0}$ & $\bar{\nu} p \rightarrow p \pi^{-} $ \\ 
%   \hline
% DCC & 6.25 & 1.53 & 1.76 & 1.21 \\
% \hline
% Hybrid & 4.82 & 1.85 &  & 1.09 \\ 
% \hline
%\end{tabular}
%    \caption{ MINERvA flux- averaged total cross section ($10^{-40} cm^{2}/nucleon $) for the single-nucleon reaction.}
 %   \label{totalxsec-table-singlenucleon}
%\end{table}

\begin{figure*}[!h]
\centering
  \includegraphics[width=\textwidth]{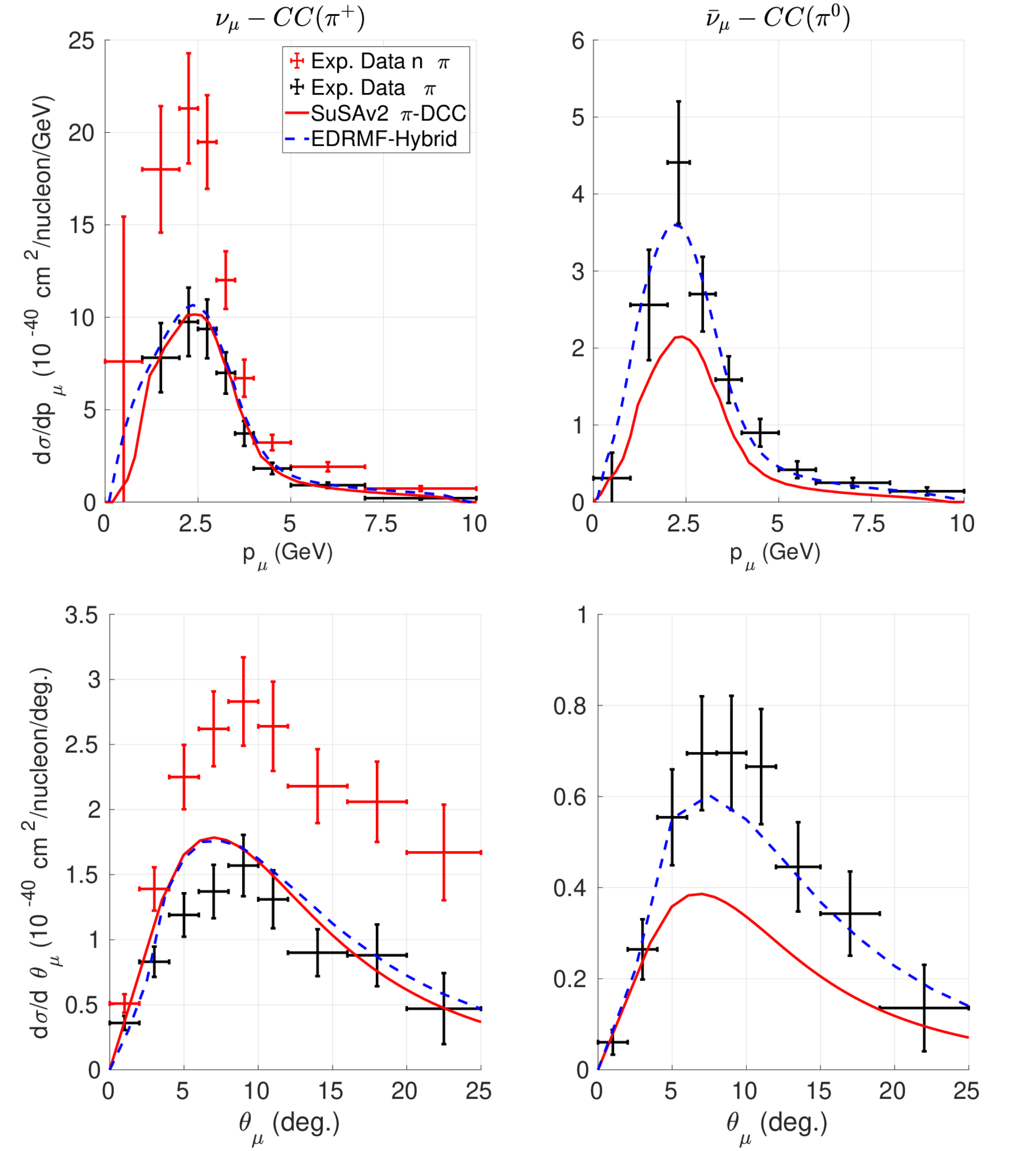}
   \caption{ MINERvA flux-averaged CC$\pi^{+}$ $\nu_{\mu}$-CH (left) and CC1$\pi^{0}$ $\bar{\nu}_{\mu}$-CH (right) differential cross sections as a function of the muon momentum (top) and the scattering angle (bottom) for multiple- (black data) and single-pion production (red data). The kinematical restrictions are (left) $W^{exp}<$ 1.4 GeV, (right) $W^{exp}<$ 1.8 GeV and 1.5 $<E_{\nu}/$GeV $<$ 10 for both cases. The experimental data points are taken from   \cite{PhysRevD.94.052005,updated1piminerva} . \label{MINERvA_PiPlus} } 
\end{figure*}

\begin{figure*}[!h]
\centering
  \includegraphics[width=\textwidth]{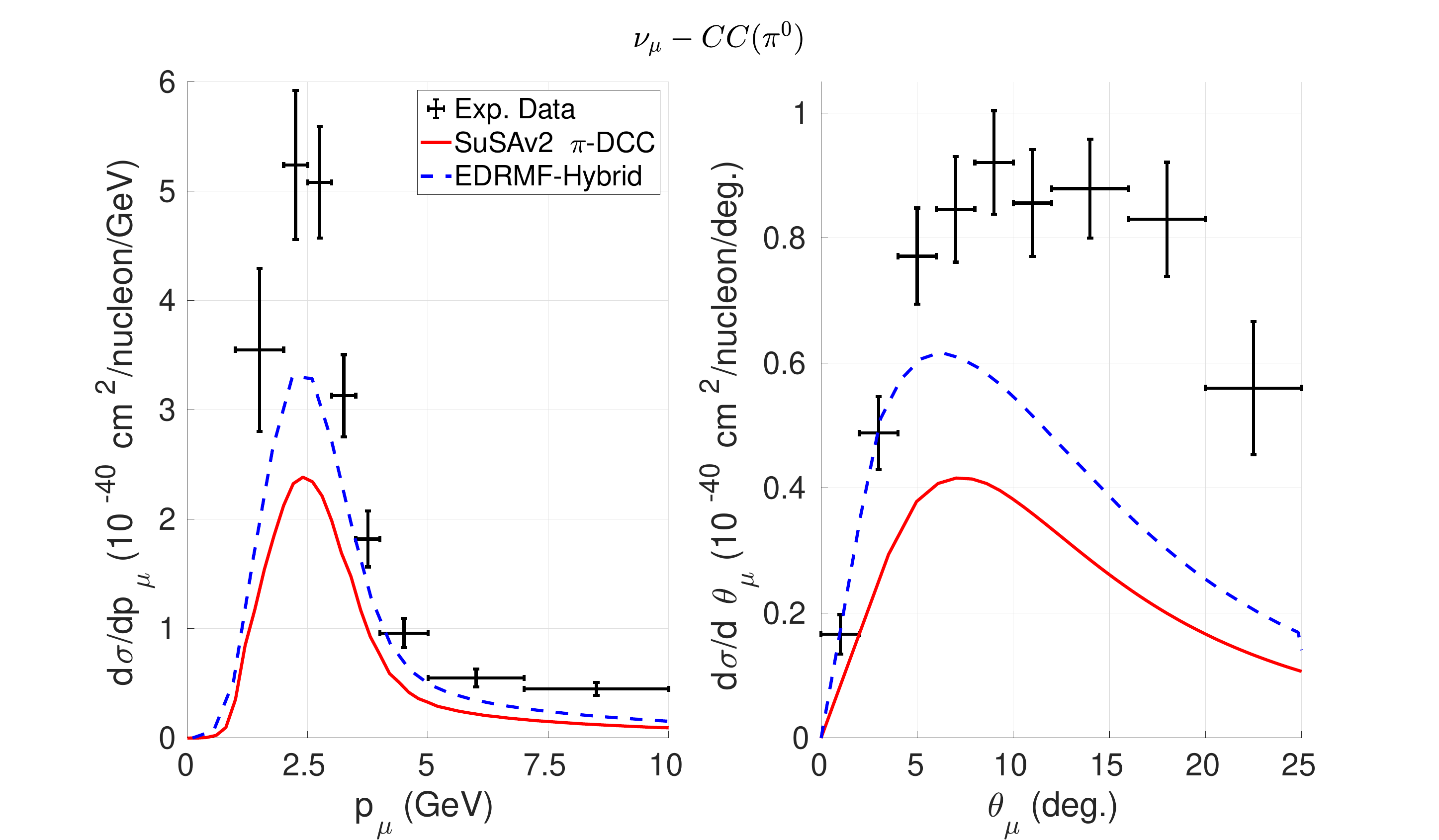}
    \caption{ MINERvA flux-averaged CC1$\pi^{0}$ $\nu_{\mu}$-CH (left) differential cross sections as a function of the muon momentum (left) and the scattering angle (right). The kinematical restrictions are 1.5 $<E_{\nu}/$GeV$<$ 20, % {\ttorange 20 or 10?} \textcolor{red}{It is 20 GeV. Each one of these curves has different restriction.}, 
$\theta_{\mu}<25$ deg. and $W^{exp}<$ 1.8 GeV. The experimental data points are taken from~\cite{PhysRevD.96.072003}. \label{MINERvA_PiZero}}
\end{figure*}

\begin{figure*}[!h]
\centering
  \includegraphics[width=\textwidth]{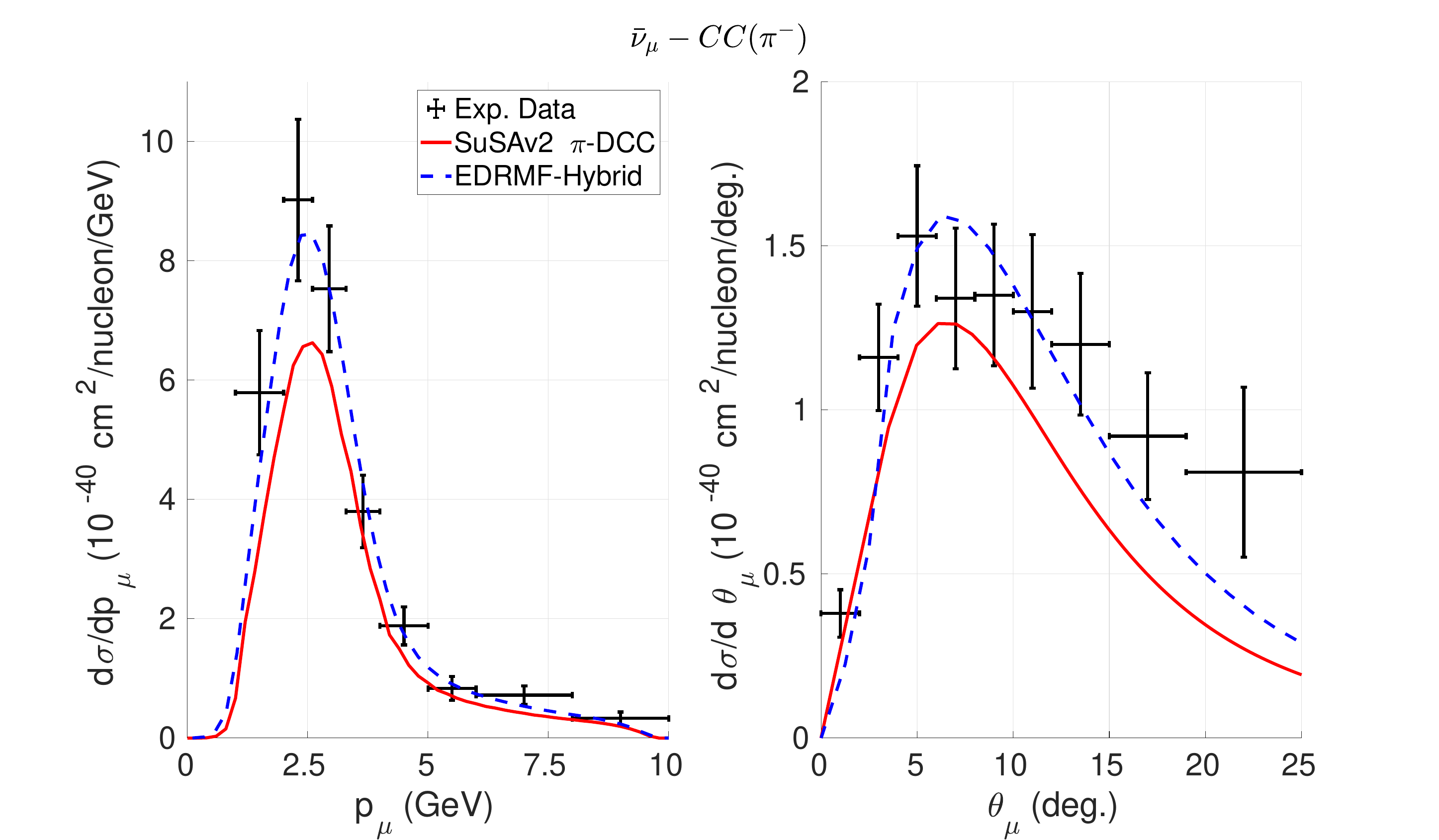}
    \caption{ MINERvA flux-averaged CC1$\pi^{-}$ $\bar{\nu}_{\mu}$-CH (left) differential cross sections as a function of the muon momentum (left) and the scattering angle (right). The kinematical restrictions are 1.5 $<E_{\nu}/$ GeV $<$ 10, $\theta_{\mu}<25$ deg. and $W^{exp}<$ 1.8 GeV. The experimental data points are taken from~\cite{PhysRevD.100.052008}.   \label{MINERvA_PiMinus}}
\end{figure*}

%{\ttorange There was no mention about Fig. 10:}
 In Fig.~\ref{comparison_proton_minerva}, the MINERvA flux-folded CC1$\pi$ single-differential cross sections in terms of the muon momentum and scattering angle for different neutrino- and antineutrino-induced pion production reactions on single-nucleons are analyzed to compare differences between DCC and Hybrid before including these approached in the corresponding nuclear models. 
 We observe larger cross sections for DCC in the $\pi^+$ case, while for the $\pi^-$ and $\pi^0$ channels the Hybrid provides larger results.
 % being the Hybrid approach for the $\pi^-$ and $\pi^0$ channels, particularly for the latter. 
 These analyses are useful to understand the following results, when the models are compared with data for pion production on the nucleus.
 % CC1$\pi$ neutrino-nucleus measurements. 
 Note that for the results shown in Fig.~\ref{comparison_proton_minerva} and in Figs.~\ref{MINERvA_PiPlus},~\ref{MINERvA_PiZero} and~\ref{MINERvA_PiMinus}, we have considered the following experimental restrictions: 
 for $\pi^+$, 1.5 $<E_{\nu}/$GeV $<$ 10 and $W^{exp}<$ 1.4 GeV; 
 for antineutrino $\pi^0$, 1.5 $<E_{\nu}/$GeV $<$ 10 and $W^{exp}<$ 1.8 GeV; 
 for neutrino $\pi^{0}$, 1.5 $<E_{\nu}/$GeV $<$ 20, $\theta_{\mu}<$ 25 deg.~and $W^{exp}<$ 1.8 GeV; 
 and for $\pi^-$, 1.5 $<E_{\nu}/$GeV $<$ 10, $\theta_{\mu}<$ 25 deg.~and $W^{exp}<$ 1.8 GeV. 
 The restrictions to relatively low values of the invariant mass imply a more prominent contribution of the $\Delta$ resonance, while the restrictions to forward angles imply focusing, on average, on smaller values of the transferred energy and momentum, where nuclear effects are more important. 

Regarding the different experimental restrictions in the scattering angle and the invariant mass applied in Fig.~\ref{comparison_proton_minerva}, the threshold in $W^{exp}<1.4$ GeV significantly reduces the cross section while the limit of 1.8 GeV barely affects this analyses at MINERvA kinematics. The restrictions in the scattering angle mainly reduce the cross section in the region of low muon momentum, and the cuts at low $E_\nu$ do not significantly affect the interpretation of the results.
 % These results are also consistent with the total cross-section analyses shown in Fig.~\ref{totalxsec} and Table~\ref{multiprogram}.

% In Figs.~\ref{MINERvA_PiPlus},~\ref{MINERvA_PiZero} and~\ref{MINERvA_PiMinus} we present the CC1$\pi$ single-differential cross sections as a function of the muon momentum and the scattering angle using the SuSAv2 $\pi$-DCC and the EDRMF-Hybrid nuclear models. The sets of plots are displayed in terms of the muon momentum and the scattering angle, respectively, which are complementary and yield the same theoretical total cross section in each case, confirming the consistency of our calculations. 

In Fig.~\ref{MINERvA_PiPlus}, we show results for CC neutrino-induced $\pi^+$ and antineutrino-induced $\pi^0$ production. For the neutrino CC1$\pi^+$ channel, two experimental datasets are shown: one corresponding to single-pion events and another that includes events with one or more pions ($n\pi$).
%{\ttorange (could we add here the inclusive-DCC results to compare with these data? Did we do it in the thesis?)} \textcolor{red}{It was not done in the thesis. It can be done, but inclusive does not distinguish between the particles in the final state. We will be including results with $\pi^{0}$ in the final state and possibly double pion, but the cut W can take care of the last one.}.  
We observe a general good agreement with data for this channel using both models, though some overestimation is obtained 
%\sout{at very forward angles} 
for scattering angles between 5 and 10 deg. 

In the antineutrino CC1$\pi^0$ channel, the SuSAv2-DCC results exhibit a similar lack of strength as observed for MiniBooNE in Fig.~\ref{MiniBooNE_PiZero}, producing the EDRMF-Hybrid approach better agreement with data. This is in accordance with the DCC vs. Hybrid differences observed in Figs.~\ref{totalxsec} and~\ref{comparison_proton_minerva} for the free-nucleon case.

In the neutrino CC1$\pi^0$ channel, Fig.~\ref{MINERvA_PiZero}, we observe an important underestimation of the data. In this case, also the EDRMF-Hybrid model underpredict the data. 
As mentioned for MiniBooNE, Fig.~\ref{MiniBooNE_PiZero}, other processes not included in this analysis, such as pion rescattering effects, could help to explain these discrepancies.  However, according to \cite{PhysRevD.97.013004}, the MINER$\nu$A cross section would not be significantly modified.
Nevertheless, it is worth mentioning that, for the antineutrino case, the EDRMF-Hybrid model produces a reasonable agreement with data at these kinematics. 
%{\ttmag A.N. (I am adding here this short discussion about this, feel free to edit:)}
As discussed in~\cite{Nikolakopoulos:2021nmb}, this is peculiar since ${}^{12}$C is a symmetric nucleus. Based on isospin symmetry, the leading order difference between the neutrino and antineutrino case is then the change of sign of the vector-axial interference contribution. The inconsistency in the description of these data may hence point to large uncertainties in the axial nucleon-to-resonance transition form factors in the second resonance region~\cite{PhysRevD.74.014009, Sato:2021pco}.  

Finally, for the antineutrino CC1$\pi^-$ channel, Fig~\ref{MINERvA_PiMinus}, we observe that the EDRMF-Hybrid agrees with the data reasonably well while SuSAv2-DCC underestimate them. It can be ascribed to the differences between DCC and Hybrid shown in Figs.~\ref{totalxsec} and~\ref{comparison_proton_minerva} for the free-nucleon case.

%{\orange In general, when comparing the ratios between DCC and Hybrid results in Fig.~\ref{comparison_proton_minerva} for scattering off single-nucleons at MINERvA kinematics with the SuSAv2-DCC and EDRMF-Hybrid ones in Figs.~\ref{MINERvA_PiPlus}-\ref{MINERvA_PiMinus}, it is observed that that the different description of nuclear dynamics in SuSAv2 and EDRMF produces an additional increase in EDRMF-Hybrid results with respect to SuSAV2-DCC ones that is not present for the single-nucleon cross sections.}
%{\ttmag A.N.: (I would write this slightly differently, because the 'increase' in EDRMF is somewhat confusing. see edit below (but you can keep what you want):)}
In general, when comparing the ratios between DCC and Hybrid results in Fig.~\ref{comparison_proton_minerva} for scattering off single-nucleons at MINERvA kinematics with the SuSAv2-DCC and EDRMF-Hybrid ones in Figs.~\ref{MINERvA_PiPlus}-\ref{MINERvA_PiMinus}, the cross section per nucleon for a bound nucleon compared to a free one is reduced in both models. This reduction is found to be  stronger in the SuSAv2 model than in the EDRMF. 

\subsubsection{T2K} \label{T2K}

\begin{figure*}[!h]
\centering
  \includegraphics[width=\textwidth]{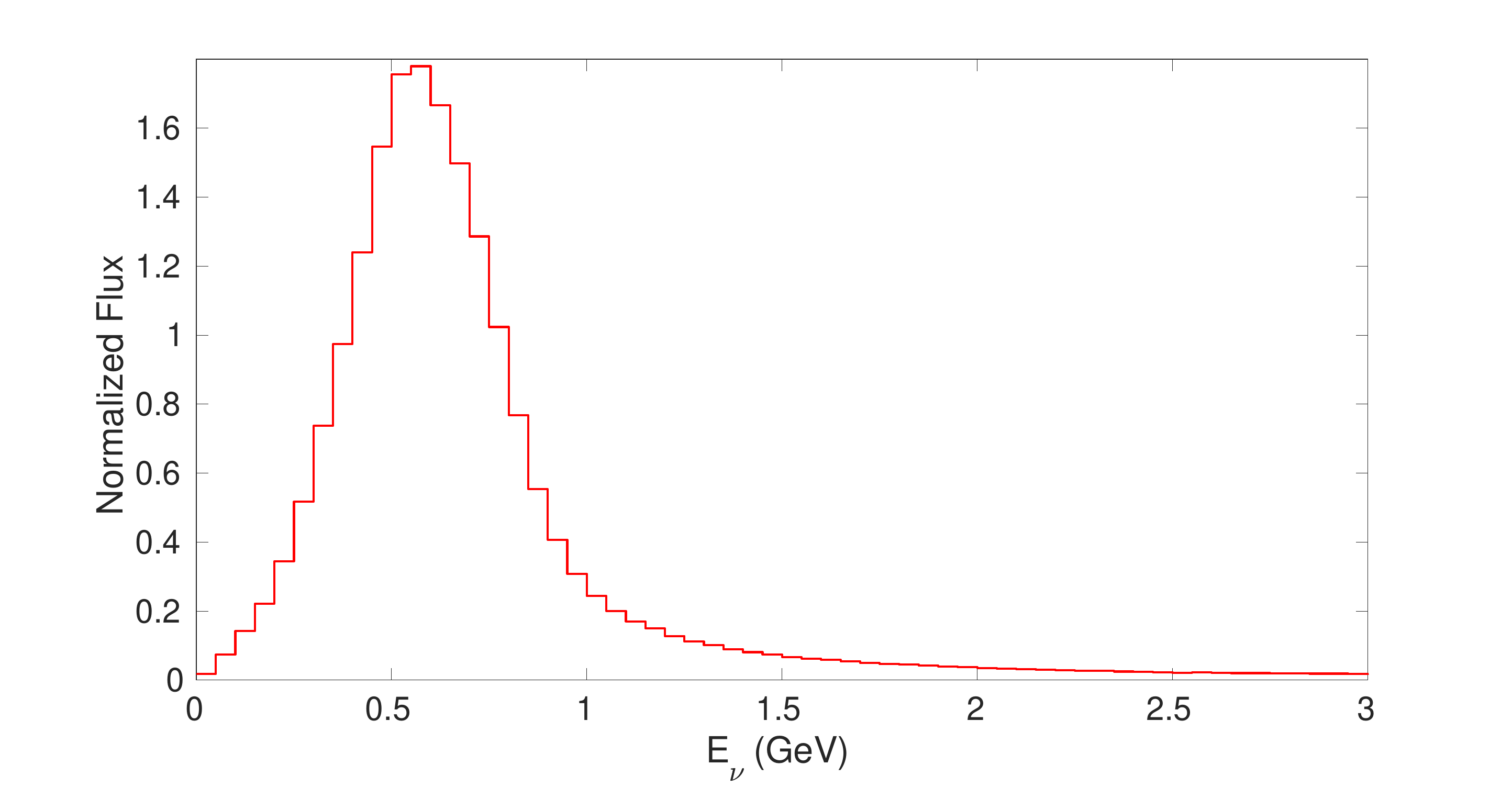}
    \caption{ Normalized neutrino flux from T2K~\cite{PhysRevD.101.012007}.
    \label{Flux_T2K}}
\end{figure*}

In this section, the results for CC1$\pi^{+}$ at T2K kinematics are shown. The target of the T2K experiment is $C_{8}H_{8}$. The normalized neutrino flux is shown in Fig.~\ref{Flux_T2K}. It peaks at around 0.6 GeV, covering an energy range similar to the MiniBooNE flux (Sect.~\ref{MiniBooNE}). In this case, the restriction $W^{exp}<2.0$ GeV is applied, so contributions from processes beyond the $\Delta$ region are not removed as was the case for the MiniBooNE data.
However, the main contribution to the $\pi^+$ production cross section remains the delta resonance, especially since the T2K flux peaks at relatively low energy.
%{\ttmag (A.N. : I rewrote the previous sentence it used to be : "the restriction $W^{exp}<2.0$ GeV is applied, so contributions from processes beyond the $\Delta$ region are expected to be larger than in the previous scenarios." In my experience this is not the case: this pion production channel is dominated by delta, and T2K has a low energy such that it definitely dominate. See for example fig 6.21 in my thesis.) }
%expected to be larger than in the previous scenarios.

\begin{figure*}[!h]
\centering
  \includegraphics[width=\textwidth]{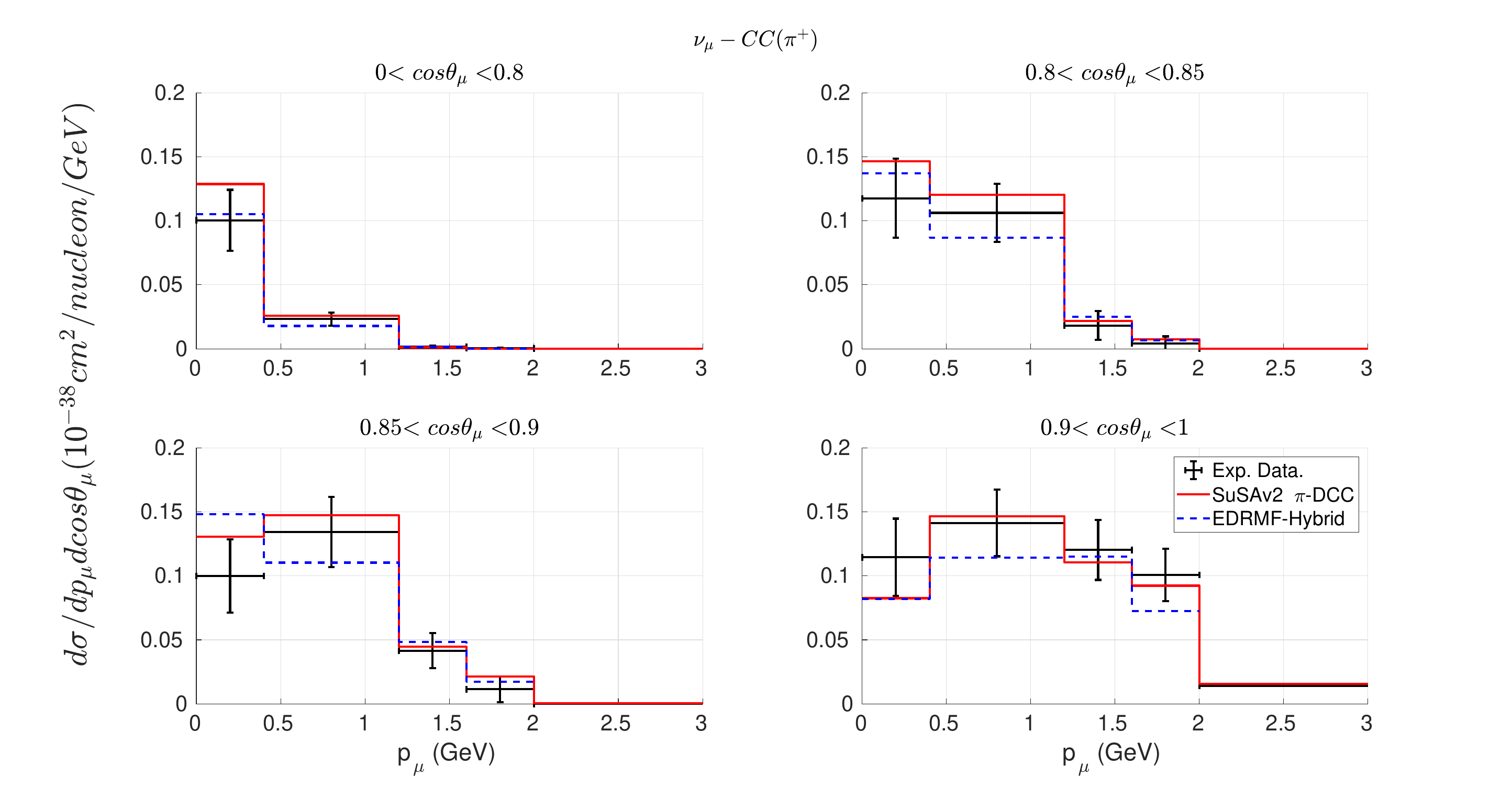}
    \caption{ T2K flux-averaged CC1$\pi^{+}$ $\nu_{\mu}$-CH differential cross sections as a function of the muon energy in bins of the cosine of the scattering angle with $W^{exp}<$ 2.0 GeV. The experimental T2K points are taken from~\cite{PhysRevD.101.012007}. 
    \label{T2K_PiPlus} }
\end{figure*}

Fig.~\ref{T2K_PiPlus} shows the flux-averaged $\nu_\mu$ CC1$ \pi^{+}$ double-differential cross section in bins of the muon scattering angle. 
% for the SuSAv2 $\pi$-DCC and EDRMF-Hybrid models in contrast to the data. 
The results are consistent with those shown for MiniBooNE. The predictions from EDRMF and SuSAv2 are close to the data and similar between them, with EDRMF producing smaller cross sections in general, which is consistent with the DCC and Hybrid differences observed in Fig.~\ref{totalxsec}. It is worth mentioning that, as in the previous cases, coherent pion production and pion rescattering effects are not taken into account by the models. %In regards to the differentiation between models, the predictions are very similar to the point that we cannot say if one model is more accurate than the other. 

%{\ttorange T2K results are available for carbon and water. Check if SuSAv2 results are for 12C or 16O. We can include both if it does not take much time.} \textcolor{red}{The results for water are really weird, probably something went wrong, but I did not have time to correct them. These results have cuts in pion momentum and pion angle that we cannot reproduce.}

%\subsubsection{MicroBooNE} \label{MicroBooNE}

%This section presents our comparison with the results for the $\nu_\mu$ flux-averaged CC1$\pi^0$ single-differential cross section measured by the MicroBooNE experiment. The target material is argon (Ar), and the neutrino flux used is shown in Fig.~\ref{Flux_Microboone}. %{\ttorange any W cut or other to mention for MicroBooNE?} \textcolor{red}{No.}

%In Fig.~\ref{MicroBooNE_PiZero}, where the SuSAv2 $\pi$-DCC is compared with the CC1$\pi^{0}$ MicoBooNE measurements, the predicted cross section underestimates the data, which is consistent with the results of the $\pi^{0}$ production shown in the previous sections for MiniBooNE and MINERvA, expecting in the same way a larger result in the case of the EDRMF-Hybrid approach.

%\textcolor{red}{Maybe comment on why the EDRMF is not included in this case??} {\ttorange I agree.}

%\begin{figure*}[!h]
%\centering
%  \includegraphics[width=\textwidth]{Flux_MicroBooNE.pdf}
%    \caption{ Normalized neutrino flux from MicroBooNE~\cite{MicroBooNE:2018efi}.
%    \label{Flux_Microboone}}
%\end{figure*}

%\begin{figure*}[!h]
%\centering
 % \includegraphics[width=\textwidth]{PionProduction_CSPiZero_MicroBooNE.pdf}
 %   \caption{  MicroBooNE flux-averaged CC1$\pi^{0}$ $\nu_{\mu}$-Ar single differential cross sections as a function of the muon momentum (left) and the cosine of the muon scattering angle (right). The experimental data points are taken from~\cite{PhysRevD.110.092014}.
%    \label{MicroBooNE_PiZero}}
%\end{figure*}

\section{Conclusions}\label{Conclusions}
%{\ttorange We could briefly mention something about EDRMF results on argon for further works if you consider it appropriate.}

%\sout{This paper presents an extensive comparison of the SuSAv2 and EDRMF approaches with neutrino- and antineutrino-nucleus CC1$\pi$ cross-section measurements, applying, respectively, two frameworks for the description of the inner structure of the nucleons in the resonance regime, namely, the DCC and Hybrid models.} 
%{\ttorange (New paragraph according to Raul's suggestion:)}
%{\ttblue (I am not sure about the `inner structure of the nucleons'. The two models, DCC and Hybrid, do more than that; they describe `microscopically' the production of a pion induced by a lepton. There are electroweak form factors in the models, which are related with the inner structure of the nucleons, but they are just another ingredient in the big pot.)}
%{\ttblue I would start differently, something like: }
This paper presents an extensive comparison of two different theoretical approaches with neutrino- and antineutrino-nucleus CC1$\pi$ cross-section measurements. The single-pion production off the nucleon is described with the DCC~\cite{nakamura_dynamical_2015,nakamura_impact_2019, DCConline,PhysRevC.67.065201,PhysRevC.88.035209} and the Hybrid~\cite{PhysRevD.95.113007,PhysRevD.97.013004} models. The DCC is incorporated into the SuSAv2-inelastic nuclear framework~\cite{PhysRevD.108.113008,gonzalez-rosa_susav2_2022,megias_charged-current_2017}, while the Hybrid is integrated into an RDWIA approach; in particular, we use the EDRMF model to treat the distortion of the final nucleon~\cite{gonzalez-jimenez_nuclear_2019}, while the pion is described as a plane wave.

%\sout{The different analysis carried out for the above-mentioned processes in combination with neutrino and antineutrino scattering off single-nucleons has yielded significant results that show a reasonable agreement for the $\pi^+$ and $\pi^-$ channels, but important differences in the $\pi^0$ channel between the DCC and Hybrid approaches.
%We also found important discrepancies between neutrino and antineutrino reactions, which could indicate for the latter that the vector-axial separation in CC1$\pi$ neutrino-nucleus interaction models could have room for improvement but also some tensions between the different experimental analyses. } 
In general, we found significant discrepancies between the two model predictions for all channels. In spite of the level of agreement observed in some figures, none of the models is able to provide a satisfactory description of all datasets.

When comparing our results with those of other theoretical approaches~\cite{PhysRevD.97.013004,PhysRevC.90.025501} and Monte Carlo event generators~\cite{PhysRevD.94.052005,PhysRevD.96.072003,PhysRevD.100.052008,PhysRevD.110.092014}, we find similar pion-production contributions to other theoretical models, while generators predict larger cross sections for $\pi^{0}$ production results. In this context, the Monte Carlo intranuclear cascade effects (mainly pion rescattering effects in the nuclear medium, leading to pion charge exchanged) could improve our agreement with data, as observed in the Supplemental Material of~\cite{PhysRevD.110.092014}, where these final-state interactions increase the CC1$\pi^0$ results. Note also the important discrepancies observed in this work for neutrino and antineutrino reactions in the $\pi^0$ channel between models and data, which may point to the importance of improvements in the modeling of axial contributions for CC1$\pi$ interactions.
%{\ttmag (A.N.: I changed part of the previous sentence, instead of 'there is room for improvements' I say : such improvements are important. (It is one of the mayor uncertainties for the modeling actually). Also I changed 'vector-axial' just to 'axial'.)}
%{\ttorange is this previous sentence still true with the recent changes in the results?} \textcolor{red}{No. It is no longer true in the case of MonteCarlo.} {\ttorange but is it still true for the models? if now everything has changed for SuSAv2 and maybe not for EDRMF-Hybrid...} \textcolor{red}{In the case of the theory is an okey comparison, the generators tend to give a much higher cross sections.} All of these approaches yield similar pion production results.

It is interesting that variations at the nucleon level, i.e., differences between DCC and Hybrid [Figs.~\ref{totalxsec} and \ref{comparison_proton_minerva}], are generally more relevant than those introduced by the nuclear model, i.e. differences due to the use of SuSAv2 or EDRMF.  A slight general increase of the EDRMF results with respect to the SuSAv2 ones has also been observed, in particular at MINERvA kinematics.
However, the ratio of cross sections per bound nucleon compared to the free nucleon is generally slightly smaller for SuSAv2 compared to the EDRMF, in particular in MINERvA kinematics.
%\sout{except at very forward angles, i.e. low energy and momentum transfer, where significant differences between EDRMF and SuSAv2 {\orange can be expected, in particular, a decrease in the EDRMF results}. The latter is connected with previous works where these effects have been analyzed in detail for the quasielastic regime~\cite{gonzalez-jimenez_constraints_2020,amaro_electron-_2020}, being related to the different treatment of low-energy nuclear effects in the SuSAv2 and RDWIA models. {\orange Apart from that, a slight general increase of the EDRMF results with respect to the SuSAv2 ones has also been observed at MINERvA kinematics (check)}. \textcolor{red}{Yes.}}

The uncertainties associated with the inelastic structure function in these models are still under investigation. Due to the limited neutrino reaction data, the axial form factor is difficult to determine accurately and may have large uncertainties. This contrasts with the vector form factors, which are well constrained by abundant electromagnetic reaction data. The impact of these uncertainties will be investigated in future works.

The limits $Q^2<3$ GeV$^2/c^2$ and $W_x<2.1$ GeV in the DCC model have little impact on the results presented here due to the experimental constraints. Nevertheless, their role could be explored in the future via detailed studies with higher-energy neutrino experiments. 
Other recent and forthcoming datasets from different experiments, such as NOvA, T2K, MINERvA, and MicroBooNE~\cite{thenovacollaboration2025measurementpi0productionbarnumu,T2K:2025smz,MINERvA:2025hzq,MicroBooNE:2025pvb}, or on different targets, such as water~\cite{PhysRevD.95.012010} or argon~\cite{PhysRevD.110.092014}, will also be analyzed in future work. 
%{\ttgreen [M: I would put the following sentence in a new paragraph. What are "These results" referred to in the first  line?]}

The results in this work indicate that the axial contribution and the modelling of nuclear effects in pion production remain uncertain and require further study. These uncertainties may affect cross-section analyses and the reconstruction methods used in Monte Carlo event generators for oscillation analyses. %As such, this work shows that the experimental analyses for pion production are not enough to completely understand these processes, being possibly a relevant source of uncertainty in oscillation analyses.

%\sout{The studies performed by these experiments and our own findings show that we do not completely understand pion production.  Despite these experiments being an important step in these analysis, these findings points to the necessity of different studies from different experiment and kinematics.}

On the other hand, the recent implementation of superscaling and RDWIA models in neutrino event generators~\cite{Dolan:2019bxf,electronsforneutrinos:2020tbf,mckean2025implementationrelativisticdistortedwave,Khachatryan2021}, such as GENIE or NEUT, mostly done for the QE and 2p2h regimes, will be extended in the future for the pion production and the full inelastic regime. This could allow us to apply intranuclear cascade effects (e.g., rescattering processes that can convert a $\pi^+$ into a $\pi^0$) to these analyses, with the possibility of having a more accurate description of final-state interaction effects, thus improving the agreement with data. The addition of other approaches for the single-nucleon inelastic structure functions, such as the MK model~\cite{kabirnezhad_single_2018} will also be considered. Finally, the use of different nuclear potentials in the RDWIA framework will also be explored in further studies in a similar way to previous works~\cite{gonzalez-jimenez_constraints_2020,PhysRevD.106.113005}, also bearing in mind more exclusive processes, i.e. cross-section measurements in terms of the kinematics of final-state pions and nucleons, which can improve nuclear model selection.

%{\ttorange I agree. This kind of paragraph can go to the Conclusions.}
%\textcolor{red}{I think this paragraph should be erased from this section. Part of the content can be used for the conclusions}
%After this analysis of CC1$\pi$ cross sections, it can be concluded that in the case of $\pi^+$ production, the SuSAv2 $\pi$-DCC model  \sout{resembles the majority of the cross section arising from the $\Delta$ resonance} {\ttorange how can we conclude that?}. In $\pi^{0/-}$ production,  the theoretical model gives half or three quarters of the remaining strength observed in the data presented throughout this chapter. When comparing these results with those of other theoretical models~\cite{PhysRevD.97.013004,Nikolakopoulos_2023,PhysRevC.90.025501} and Monte Carlo generators~\cite{PhysRevD.94.052005,PhysRevD.96.072003,PhysRevD.100.052008,PhysRevD.110.092014}, we find good agreement. All of these approaches yield similar pion production results. In this regard, the superscaling model is consistent with other theoretical \sout{and simulation-based} predictions.

\begin{acknowledgments}
This work has received funding from the European Union's Horizon research and innovation programme under the Marie Skłodowska-Curie Actions (MSCA) HORIZON-MSCA-2023-SE-01-01 - MSCA Staff Exchanges 2023, Grant Agreement ID: 101183137 (JENNIFER3 project), and is partially supported by the Spanish Ministerio de Ciencia, Innovaci\'on y Universidades and ERDF (European Regional Development Fund) under contract PID2023-146401NB-I00, by the Junta de Andalucia (grants No. FQM160 and SOMM17/6105/UGR), by University of Tokyo ICRR’s Inter-University Research Program FY2025 and FY2026 (Refs. 2025i-J-001 and 2026i-J-001), by the INFN under project Iniziativa Specifica NucSys and the University of Turin under Project BARM-RILO-24 (M.B.B.). J.G.R. was supported by a Contract PIF VI-PPITUS 2020 from the University of Seville (Plan Propio de Investigaci\'on y Transferencia).
R.G.-J. was supported by projects PID2021-127098NA-I00 and RYC2022-035203-I funded by MCIN/AEI/10.13039/501100011033/FEDER and FSE+, UE;
and by ``Ayudas para Atracción de Investigadores con Alto Potencial-modalidad A'' funded by VII PPIT-US. %\textcolor{red}{Jesus: No longer supported by the contract, but the research was done under the contract. Should be include it?}. {\ttorange Yes, we can include it.}
A.N. is supported by the Neutrino Theory Network (NTN) under Award Number DEAC02-07CH1135.
\end{acknowledgments}

\bibliography{CCRES.bib}

\end{document}